\documentclass{aa}

\def\kms{{\rm km\,s^{-1}}}

\def\mVR{$\langle V_R \rangle$}

\newcommand{\mycomment}[1]{}

\let\oldsim\sim 
\renewcommand{\sim}{{\oldsim}}

\usepackage[normalem]{ulem}
\usepackage{graphicx}
\usepackage{txfonts}
\usepackage{amsmath}
\usepackage{yhmath}
\usepackage{algorithm,algorithmic}
\usepackage{xcolor}
\usepackage{bm}
\usepackage{multirow}
\usepackage{svg}
\usepackage{makecell}

\defcitealias{antoja2022tidal}{A22}

\begin{document}

   \title{Dynamics of tidal spiral arms: Machine learning-assisted identification of equations and application to the Milky Way}
   \titlerunning{Dynamics of tidal spiral arms}
   
   \author{Marcel Bernet\inst{1,2,3}
            \and Pau Ramos \inst{4}
            \and Teresa Antoja \inst{1,2,3}
            \and Adrian Price-Whelan \inst{5}
            \and \\ Steven L. Brunton \inst{6,7}
            \and Tetsuro Asano \inst{1,2,3}
            \and Alexandra Girón-Soto \inst{1,2,3}
    }
    
    \institute{Departament de Física Qu\`antica i Astrof\'isica (FQA), Universitat de Barcelona (UB),  c. Mart\'i i Franqu\`es, 1, 08028 Barcelona, Spain \email{mbernet@fqa.ub.edu}
    \and{Institut de Ci\`encies del Cosmos (ICCUB), Universitat de Barcelona (UB), c. Mart\'i i Franqu\`es, 1, 08028 Barcelona, Spain}
    \and{Institut d'Estudis Espacials de Catalunya (IEEC), Edifici RDIT, Campus UPC, 08860 Castelldefels (Barcelona), Spain}
    \and{National Astronomical Observatory of Japan, Mitaka-shi, Tokyo 181-8588, Japan}
    \and{Center for Computational Astrophysics, Flatiron Institute, 162 Fifth Ave, New York, NY 10010, USA}
    \and{Department of Mechanical Engineering, University of Washington, Seattle, Washington 98195, USA}
    \and{AI Institute in Dynamic Systems, University of Washington, Seattle, WA 98195, USA}
    }
    \date{Received YYY; accepted XXX}

 
  \abstract
    {Understanding the spiral arms of the Milky Way (MW) remains a key open question in galactic dynamics. Tidal perturbations, such as the recent passage of the Sagittarius dwarf galaxy (Sgr), could play a significant role in exciting them.}
    {We aim to analytically characterize the dynamics of tidally induced spiral arms, including their phase-space signatures.}
    {We ran idealized test-particle simulations resembling impulsive satellite impacts, and used the Sparse Identification of Non-linear Dynamics (SINDy) method to infer their governing Partial Differential Equations (PDEs). We validated the method with analytical derivations and a realistic $N$-body simulation of a MW-Sgr encounter analogue.}
    {For small perturbations, a linear system of equations was recovered with SINDy, consistent with predictions from linearised collisionless dynamics. In this case, two distinct waves wrapping at pattern speeds $\Omega \pm \kappa/m$ emerge, where $\Omega$ and $\kappa$ are the azimuthal and epicyclic frequencies, and $m$ is the azimuthal mode number.
    For large impacts, we empirically discovered a non-linear system of equations, representing a novel formulation for the dynamics of tidally induced spiral arms. 
    For both cases, these equations describe wave properties like amplitude and pattern speed, and their shape and temporal evolution in different phase-space projections.
    In the realistic simulations, we recovered the same equation. However, the fit is sub-optimal, pointing to missing terms in our analysis, such as velocity dispersion and self-gravity.
    We fit the \emph{Gaia} $L_Z-$\mVR{ }waves with the linear model, providing a reasonable fit and plausible parameters for the Sgr passage. However, the predicted amplitude ratio of the two waves is inconsistent with observations, supporting a more complex origin for this feature (e.g. multiple passages, bar, spiral arms).}
    {We merge data-driven discovery with theory to create simple, accurate models of tidal spiral arms that match simulations and provide a simple tool to fit \emph{Gaia} and external galaxy data. 
    This methodology could be extended to model complex phenomena like self-gravity and dynamical friction.}


   \maketitle

%
%

\section{Introduction}


Understanding the dynamics of the spiral arms of the Milky Way (MW) remains a major challenge in galactic astronomy \citep[][]{dobbs2014baba,sellwood2022review}. Theoretical models range from density wave theories \citep{linshu1964,kalnajs1973density}, to transient and recurrent, swing‑amplified instabilities \citep{goldreich1965swing,julian1966swing}, and self‑excited spiral modes \citep{sellwood1984instabilities}, with varying consequences on the spiral pattern speed, pitch angle, and bar–spiral coupling \citep{lyndenbell1972barspiral,sanders1976barspiral,sellwood2022review}. In the MW, although spiral arm segments can be traced by young stars and gas \citep{georgelin1976spiralarm, drimmel2001spiralarm,reid2009spiralarms}, their number, longevity, and origin are still debated.

The \emph{Gaia} mission \citep{gaia2016} has revealed complex kinematic substructures across various phase space projections, $R-V_\phi$ plane \citep[e.g.][]{antoja2018phasespiral, kawata2018ridges,fragkoudi2019ridges, ramos2018ridges, bernet2024movinggroups}, wave-like patterns in $R-V_R$ and $L_Z-V_R$ \citep[e.g.][]{friske2019wave, eilers2020strength, antoja2022tidal, hunt2024rvr_spiral}, thin arches in $V_R-V_\phi$ \citep[e.g.][]{katz2018dynamics, bernet2022movinggroups, bernet2024movinggroups}, and a vertical phase spiral in $Z-V_Z$ \citep[e.g.][]{antoja2018phasespiral,laporte2019phasespiral, blandhawthorn2021phasespiral, hunt2021spiral, antoja2023phasespiral, darragh-ford2023phasespiral}.
Interpreting the origin and evolution of these global disequilibrium features requires theoretical modelling that accounts for the effects of the bar, spiral arms, and perturbations with external galaxies.

In \citeauthor{antoja2022tidal}\,(\citeyear{antoja2022tidal}, hereafter \citetalias{antoja2022tidal}), motivated by the evidence of ongoing vertical phase mixing in the disc \citep{antoja2018phasespiral} probably due to the interaction with the Sagittarius (Sgr) dwarf galaxy \citep{ibata1994sgr}, we developed a framework to interpret some of the planar features observed in the MW disc in terms of tidally induced spiral arms. By combining analytical models with test particle simulations and an $N$-body model of the Sgr–MW interaction \citep{laporte2018sims}, we showed how a tidal impact produces spiral arms that phase-wrap over time and manifest as coherent ridges in $R-V_\phi$ and  waves in $L_z-V_R$. However, \citetalias{antoja2022tidal} was limited to deriving only the spiral arm loci over time, as the full dynamics captured in simulations can be remarkably challenging to model analytically.

While a deep foundation in theoretical and analytical models has historically shaped our understanding of galactic dynamics \citep[e.g.][]{binney2008bt}, the explosion of high-precision data and modern simulations is transforming the field. This presents an opportunity for data-driven approaches, where we can use the data itself to empirically discover the underlying dynamical principles. Inferring physical laws from data is a long-standing practice in astronomy. Centuries ago, Kepler employed symbolic regression to derive his empirical laws of planetary motion \citep{kepler1609}. More recently, \citet{tenachi2023sr,tenachi2024sr} have introduced a general deep learning framework for extracting analytical governing laws from simulated and/or observed data, which they successfully applied to derive an analytic galaxy potential directly from simulated stellar orbits.

Beyond these examples in astronomy, significant advancements have been made in data-driven methods for discovering the governing equations of complex dynamical systems across various scientific domains. A pioneering approach is SINDy (Sparse Identification of Nonlinear Dynamics), introduced by \citet{brunton2016sindy}, which uses sparse regression to identify the governing ordinary differential equations (ODEs) of a dynamical system directly from time-series data. This idea was subsequently extended to partial differential equations (PDEs) by \citet[][]{rudy2017sindypde} and \citet{schaeffer2017pde}, utilizing spatial-temporal data. Techniques such as the integral or weak formulation \citep{schaeffer2017integral, reinbold2020weak} have further refined these methods, improving robustness to noise and handling high-order derivatives. These advances have enabled the data-driven discovery (or rediscovery) of governing PDEs in various physical domains, including geophysical fluids \citep{zanna2020data}, plasma physics \citep{alves2022rediscovered}, active matter \citep{supekar2023learning}, and turbulence closure modelling \citep{beetham2020formulating,schmelzer2020discovery,zanna2020data,beetham2021sparse}. A comprehensive review of these methods is provided by \citet{north2023sindyreview}. Other related techniques for discovering dynamical systems, particularly for ordinary differential equations, include symbolic regression through genetic programming \citep{Bongard2007pnas,Schmidt2009science,cranmer2023interpretable}.


In this work, we use a data-driven dynamical discovery approach to understand tidally induced spiral arm dynamics. We build idealised test-particle simulations (Sect.\,\ref{sect:simulations}) and use SINDy to infer their governing equations (Sect.\,\ref{sect:SINDy}).
This approach reveals two families of PDEs that govern spiral wave evolution. For small perturbations, we recovered a linear system of equations, consistent with predictions from established linearised collisionless theory. For large perturbations with specific profiles, SINDy empirically identified a non-linear system of equations. Notably, this PDE itself constitutes a novel mathematical formulation for the dynamics of tidally induced spiral arms. Guided by these empirical findings, we analytically derive these same equations from first principles (Sect.\,\ref{sect:analytic_derivation}), recovering closed expressions for their coefficients and obtaining explicit solutions that describe wave properties like amplitude and pattern speed, and their shape and temporal evolution in different phase-space projections. This framework is validated against realistic N-body simulations (Sect.\,\ref{sect:n-body}). Finally, we use these analytical solutions to test whether our simple models can reproduce the characteristics of the $L_Z-$\mVR{ }wave seen in the \emph{Gaia} data (Sect.\,\ref{sect:gaia_wave}).
Our work demonstrates how data-driven methods can lead to the discovery of simple governing equations that can be solved analytically and provide interpretable, predictive models for key kinematic features observed in the MW and other galaxies perturbed by internal or external forces.

%
%

\section{Test particle simulations for tidal impacts} \label{sect:simulations}

This section describes our 2D test-particle simulations. We run them with \texttt{Agama} \citep{vasiliev2019agama}. We used the classical \texttt{MWPotential2014} \citep{bovy2015galpy}, consisting of a \citet{miyamotonagai1975disk} disk, a NFW \citep{navarro1996nfw} halo and a \citet{dehnen1993bulgepot} bulge. 
The dynamical features we study are present in any reasonable axisymmetric potential (logarithmic, Plummer, etc), and, thus, our results do not depend on this choice.
We initialize all test particles on perfectly circular orbits, with zero velocity dispersion. Thus, each particle satisfies:
\begin{equation}\label{eq:ics}
    V_R = 0\,\text{km\,s}^{-1}\quad V_\phi = V_c,
\end{equation}
where $V_c$ is the circular velocity curve of the assumed model. We sampled $10^6$ particles using a uniform distribution in $R$ to increase the signal in the outer disc.

Once these initial conditions (ICs) are set up, we induce a perturbation to the system. In this section, we apply three types of perturbations to these ICs: a distant and small impact (Sect.\,\ref{sect:m2_kick}), a generalized small impact (Sect.\,\ref{sect:generalized_kick}), and a large impact (Sect.\,\ref{sect:large_impact_sims}). Each perturbation generates a distinctive kinematic signal that we aim to understand in this work.

\subsection{Distant and small impacts}\label{sect:m2_kick}

\begin{figure*}
    \centering
    \includegraphics[width=.99\linewidth]{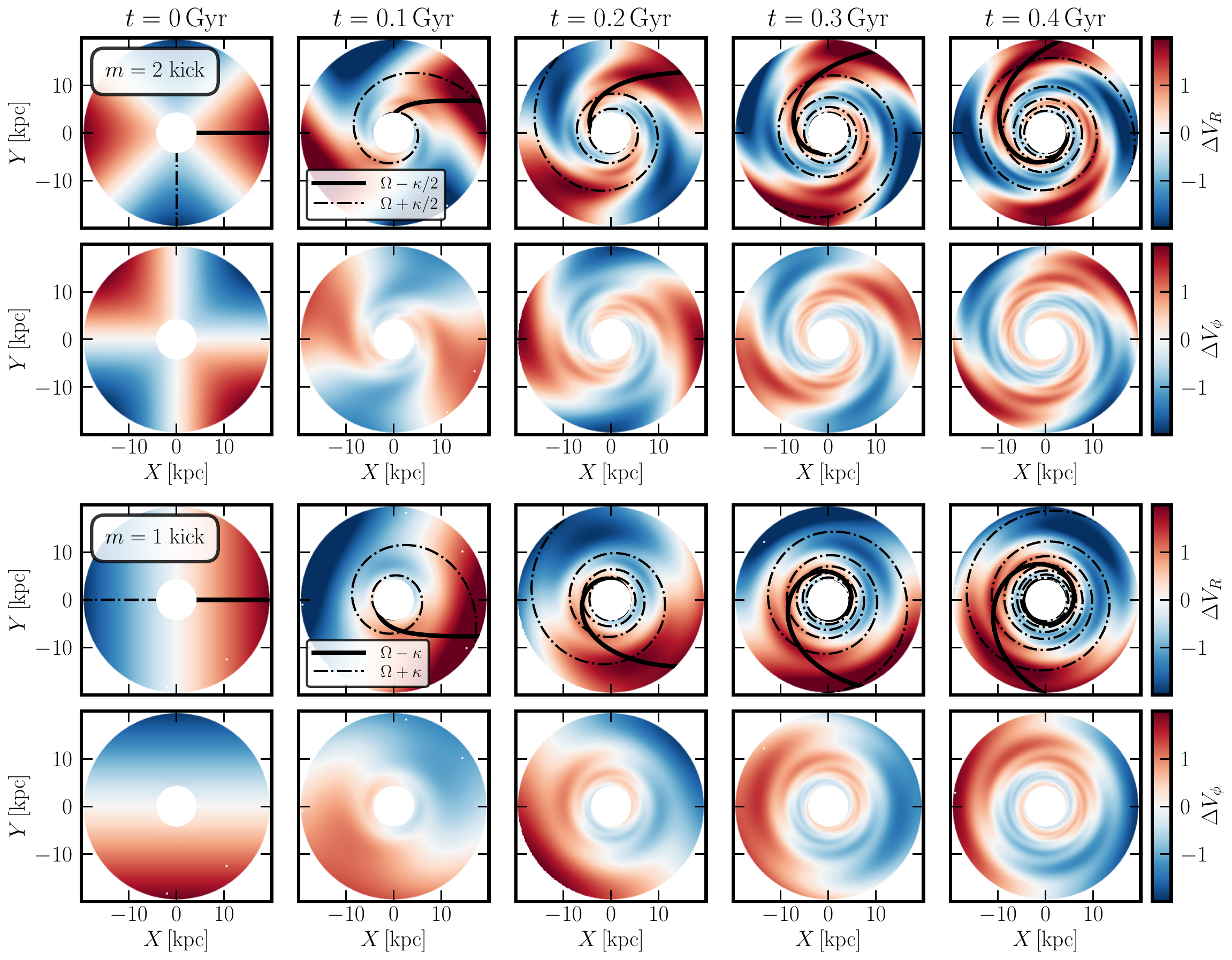}
    \caption{
        Kinematic signature evolution of the $m=2$ (upper panels) and $m=1$ (lower panels) kinematic kicks. We show snapshots of the velocity perturbation fields at five different times ($t = 0, 0.1, 0.2, 0.3, \text{ and } 0.4\,$Gyr). For each kick mode, the top row displays the radial velocity perturbation, $\Delta V_R$, while the bottom row shows the azimuthal velocity perturbation, $\Delta V_\phi$. The solid and dash-dotted curves overlaid on the $\Delta V_R$ panels indicate the theoretical spiral loci wrapping at pattern speeds $\Omega - \kappa/m$ and $\Omega + \kappa/m$, respectively. The panels demonstrate the clear appearance of two spiral wave patterns for both $m=2$ and $m=1$ modes, with the theoretical loci accurately matching the observed evolution of the waves in the velocity fields.
    }
    \label{fig:velocity_maps}
\end{figure*}

To simulate a weak and distant impulsive encounter, we employ the impulse approximation (the perturber passes quickly) and the tidal approximation (the perturber is distant). Under these strong assumptions, the complex, time-evolving gravitational interaction is simplified to a single, instantaneous velocity kick. For this work, we do not model a specific encounter orbit but instead adopt the following canonical form \citep[e.g.][]{aguilar1985tidal,struck2011tidal}, that matches, to first order, the kinematic response of realistic $N$-body models
\citepalias{antoja2022tidal}:
\begin{equation}\label{eq:m2_kick}
    \Delta V_R = \epsilon R \cos(2\phi), \quad \Delta V_\phi = -\epsilon R \sin(2\phi),
\end{equation}
which are added to the ICs defined in Eq.\,\ref{eq:ics}. The direction of closest approach is arbitrarily chosen to be $\phi=0$ (equivalent to $\phi=\pi$), resulting in the maximum kick in $V_R$ being at the same azimuth. The scale parameter $\epsilon$, that has units of km\,s$^{-1}$\,kpc$^{-1}$, sets the amplitude of this initial velocity kick. For a quick flyby by a mass $M_p$ at closest distance $r_p$ with velocity $v_p$, $\epsilon$ can be estimated as $\epsilon \approx 2\,G\,M_p / (r_p^2\,v_p)$ \citep[Sect.\,8.2.1. In][]{binney2008bt}. In this section, we assume a small impact, which we define as a perturbation where the response of the system is linear. This requires the velocity kick to be much smaller than the circular velocity ($\Delta V_R, \Delta V_\phi \ll V_c$) and, therefore, a small $\epsilon$. For this work, we quantitatively consider the small impact regime to correspond to kick amplitudes of $\epsilon \lesssim 0.1$\,km\,s$^{-1}$\,kpc$^{-1}$, which leads to a velocity kick $\lesssim 10$\,km\,s$^{-1}$ at the studied radii. The two-fold symmetry\footnote{While Eq.\,\ref{eq:m2_kick} follows most directly for a perturber moving in the disc plane, we note that inclined or polar passages produce essentially the same two‑fold ($m=2$) pattern, with only order‑unity differences in amplitude.} in the radial ($\Delta V_R$) and azimuthal ($\Delta V_\phi$) velocity fields can be seen in the leftmost panels of the upper rows in Fig.\,\ref{fig:velocity_maps}.

After evolving these initial conditions for $0.4\,$Gyr, two distinct wave modes emerge in the velocity field. The primary wave is seen mostly in the wrapping of the velocity patterns in a spiral shape (red-blue) of large scale. This leads to a density enhancement in regions of negative $V_R$ that constitute the spiral arms \citepalias[see Fig.\,4 in][]{antoja2022tidal}. The secondary wave is fainter and wraps faster, so it is seen as smaller scale oscillations on top of the main pattern. Further details and visualizations of the resulting 1D radial profiles, illustrating the contributions of the primary and secondary waves, are presented in Appendix \ref{app:radial_profiles}.
We see that the pattern speed of the primary spiral wave is $\Omega - \kappa/2$, which is widely known \citep[e.g.][]{kalnajs1973density,oh2008tidal,oh2015tidal,struck2011tidal,semczuk2017tidal}, although the presence of a bar or self-gravity can modify this pattern speed, up to $\Omega - \kappa/4$ \citep[e.g.][]{oh2015tidal,pettitt2018tidalsg}. On the other hand, in \citetalias{antoja2022tidal}, we already noted the presence of the secondary wave but we did not characterise it. 
We see here that it wraps as $\Omega + \kappa/2$. In Sect.\,\ref{sect:analytic_derivation} we give the analytical explanation for its appearance and its wrapping frequency.

\subsection{Generalised kinematic kick}\label{sect:generalized_kick}

The framework assumes an $m=2$ kinematic kick, but we can extend this formulation to include other perturbation shapes, accounting for more complex patterns with multiple $m$ modes from more realistic galactic encounters and halo misalignments \citep[e.g.][]{chakrabarti2009tidal,ghosh2022lopsidedness}. Here, we apply idealized $m$-mode kinematic kicks as a controlled experiment to isolate the dynamical response to specific shapes of the ICs. Specifically, we can generalise Eq.\,\ref{eq:m2_kick} to 
\begin{equation}\label{eq:m_kick}
    \Delta V_R = \sum_m \epsilon_m R \cos(m\phi), \quad \Delta V_\phi = -\sum_m \epsilon_m R \sin(m\phi).
\end{equation}
In the bottom group of Fig.\,\ref{fig:velocity_maps}, we show the velocity maps for an $m=1$ kick.  
In this case, the two waves wrap at frequencies of $\Omega_{\text{sp}} = \Omega \pm \kappa$. In general, for every $m$-shaped kick we include, two spiral patterns emerge, wrapping according to:
\begin{equation}\label{eq:empirical_omk}
    \Omega_{\text{sp}} = \Omega \pm \kappa/m.
\end{equation}
The basic intuition is as follows: each star circles around its reference (guiding) orbit at an angular rate $\Omega$ and, in addition, executes small radial (epicyclic) oscillations at a frequency $\kappa$. 
A single star completes one full radial oscillation—reaching repeatedly its apocentre—at this epicyclic frequency $\kappa$. 
However, the imposed $m$-fold symmetry in the initial perturbation sets the number of apocentres in a given radius. Instead of observing one apocentre per full $\kappa$ epicycle, the symmetry causes the collective apocentre pattern to repeat every $\kappa/m$. Therefore, between two consecutive apocentres of a single star, one observes additional apocentres from other stars — a total of $m$, spread evenly across the circle. This sets the precession rate, creating the pattern described in Eq.\,\ref{eq:empirical_omk}.

At a fixed radius, the amplitudes of the primary wave ($\Omega-\kappa/m$) and the secondary wave ($\Omega+\kappa/m$) remain constant in time and are independent of the azimuthal number $m$. Their relative strength, however, varies with radius and with the choice of gravitational potential. In our fiducial model we have measured the relative amplitude of both waves by computing the 2D Fourier amplitude in $\phi-t$ maps. The ratio of amplitudes between both waves stays nearly constant throughout the disk, taking a value very close to
\begin{equation}\label{eq:amplitudes_empirical}
    \frac{\mathcal{A}_{\Omega-\kappa/m}}{\mathcal{A}_{\Omega+\kappa/m}} \approx 5.2,
\end{equation}
where $\mathcal{A}$ represents the amplitude of a wave. In Section\,\ref{sect:analytic_derivation}, we recover this value from first principles.

\subsection{Single-wave large impacts}\label{sect:large_impact_sims}

\begin{figure*}
    \centering
    \includegraphics[width=.94\linewidth]{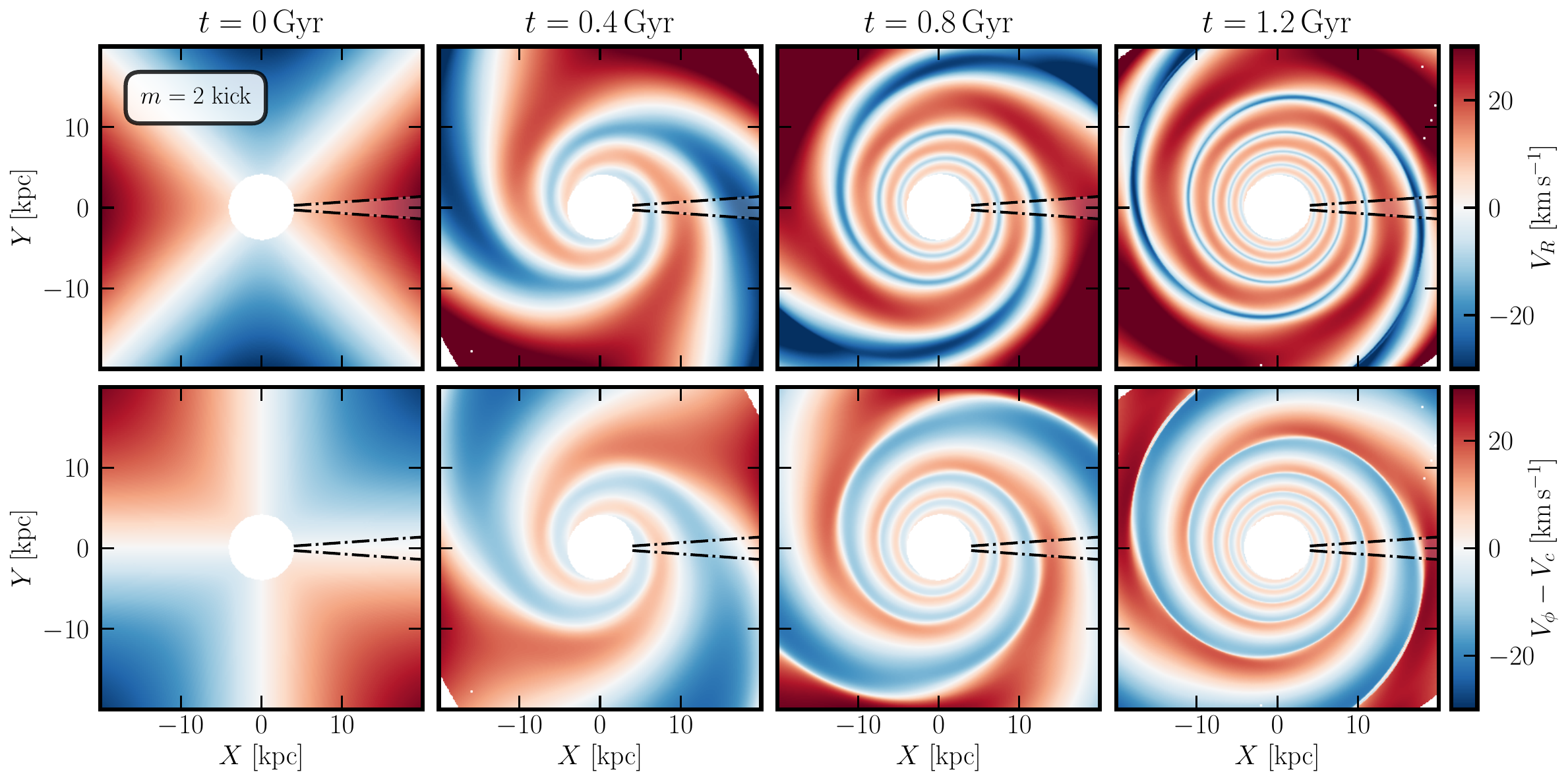}
    \caption{
        Evolution of the test-particle system after a large, $m=2$ impact, designed with a $\gamma$ correction (Eq.\,\ref{eq:m_kick_large}) to excite a single spiral pattern. These panels show snapshots of the velocity field at four different times ($t=0$, $0.4$, $0.8$, and $1.2\,$Gyr). The top row shows the radial velocity $V_R$, and the bottom row shows the residual azimuthal velocity, $\Delta V_\phi = V_\phi - V_c$. The spiral pattern winds up over time at the expected rate $\Omega - \kappa/m$. The large amplitude of the perturbation introduces non-linear effects, causing the regions of negative $V_R$ (blue) to become more concentrated into arcs, and producing sharp sign changes in $\Delta V_\phi$. One-dimensional radial profiles extracted from these maps along $\phi=0^\circ$ (dash-dot regions) are presented in Figure \ref{fig:velocity_maps_large_r}.
    }
    \label{fig:velocity_maps_large_xy}
\end{figure*}

\begin{figure*}
    \centering
    \includegraphics[width=.94\linewidth]{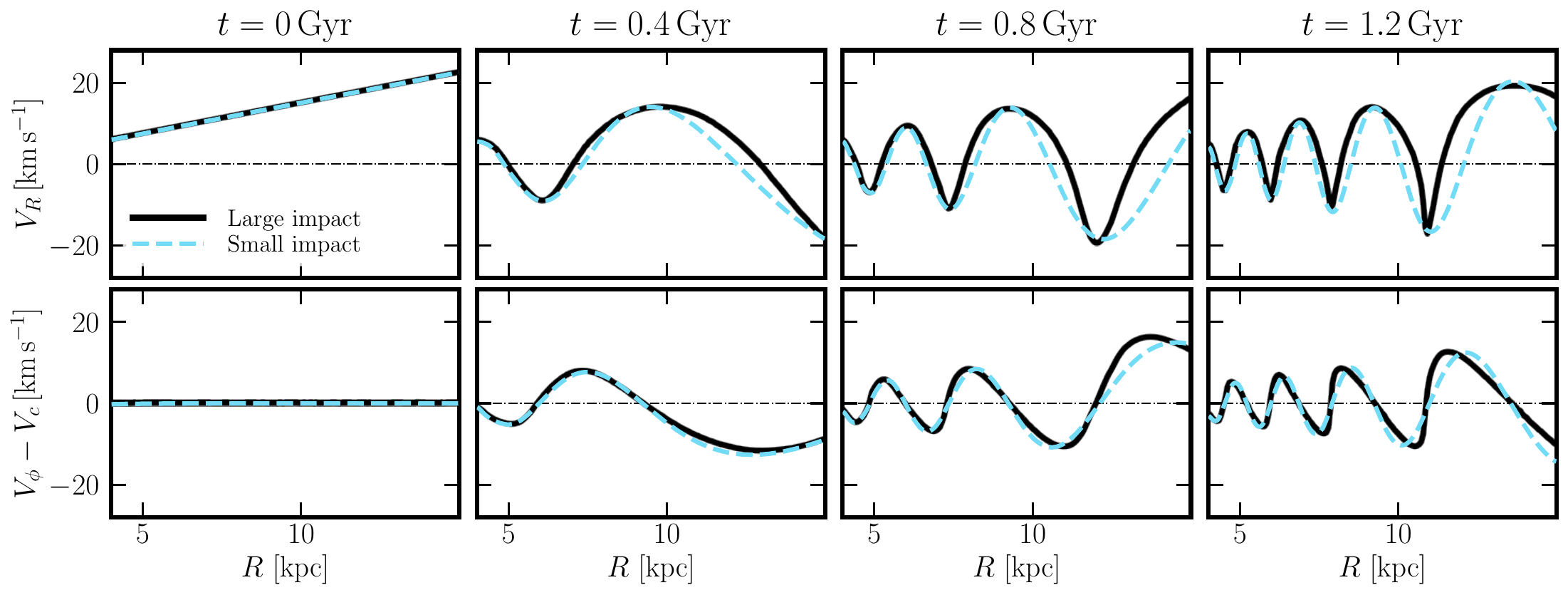}
    \caption{
        One-dimensional radial profiles of the velocity perturbations resulting from the large, $m=2$ impact simulation shown in Figure \ref{fig:velocity_maps_large_xy}. These profiles show $V_R$ (top row) and $\Delta V_\phi$ (bottom row) as a function of radius $R$, at the same four times ($t=0$, $0.4$, $0.8$, and $1.2\,$Gyr). The profiles are averaged on small radial bins at fixed azimuth ($\phi=0^\circ$, corresponding to the dashed-dot regions in Figure \ref{fig:velocity_maps_large_xy}). The solid black lines represent the results from the large impact simulation. The dashed blue lines show the velocity profiles resulting from the dominant $\Omega - \kappa/m$ wave (specifically the $\Omega - \kappa/2$ wave for $m=2$) in a small impact simulation (extracted from the blue curves in Fig.\,\ref{fig:small_1d}), scaled up for comparison. These black 1D profiles reveal the characteristic triangular wave shape in $V_R$ and the sawtooth pattern in $\Delta V_\phi$, which are key signatures of the non-linear velocity structures generated by the large initial perturbation.
    }
    \label{fig:velocity_maps_large_r}
\end{figure*}

While the generalized kinematic kick framework in Sect.\,\ref{sect:generalized_kick} assumes small perturbations that excite two modes (as shown in Sect.\,\ref{sect:m2_kick} and derived analytically in Sect.\,\ref{sect:analytic_derivation}), we now explore regimes where the velocity kick amplitude $D$ (analogous to $\epsilon$, but not restricted to small values) becomes larger ($\gtrsim 10$\,km\,s$^{-1}$ at the studied radii), pushing the system into a non-linear response regime. For these large impact simulations, we only study the dynamics of the slow wave, which has a larger amplitude. The subdominant wave phase-mixes much faster and its long-term contribution is minor \citep{toomre1969phasemixing}. This simplifies the dynamical system and facilitates our SINDy analysis. To achieve this, we introduce a specific velocity kick designed to preferentially excite one spiral mode over the other, using a $\gamma$-dependent scaling between its radial and azimuthal components: 
\begin{equation}\label{eq:m_kick_large}
\Delta V_R = D R \cos(m\phi), \quad \Delta V_\phi = -\frac{D}{\gamma} R \sin(m\phi),
\end{equation}
where $\gamma \equiv 2\Omega/\kappa$ is derived from the potential. Physically, this $\gamma$-dependent correction balances the two effective restoring coefficients in a disk ($\kappa$ for radial oscillations and $2\Omega$ for azimuthal ones) so that all stars have the same epicyclic amplitude in the radial and azimuthal direction and uniform phase distribution \citep[Chapter 3.2.3. in][]{binney2008bt}. The resulting single-wave pattern (Fig.\,\ref{fig:velocity_maps_large_xy}) is an intended outcome, designed to facilitate the SINDy discovery (Sect.\,\ref{sect:sindy_large}).

In Fig.\,\ref{fig:velocity_maps_large_xy} we show the two‐dimensional evolution of the system following this impact. As a result of the $\gamma$ correction, only a single $m$-armed spiral pattern emerges, which winds up at the expected rate $\Omega - \kappa/m$. However, the nonlinearity introduced by the large $D$ amplitude generates a non trivial pattern: regions of negative $V_R$ become progressively confined into tighter arcs, while the $V_\phi$ map develops sharp sign changes.

We can intuitively explain these non-linear wrapping patterns. At a given radius $R$, the minimum $V_\phi$ corresponds to the apocentre of an inner orbit with guiding radius $R_{\text{g, in}} < R$. As we discussed in the previous section, the pattern speed of these apocentres locus is $\Omega(R_{\text{g, in}}) - \kappa(R_{\text{g, in}})/m$, and this happens to be larger than $\Omega(R) - \kappa(R)/m$. The opposite argument can be done for the maximum $V_\phi$, which correspond to the pericentre of an outer orbit. Therefore, at a given $R$, the points with minimum (maximum) $V_\phi$ wrap faster (slower) than the rest, producing these sharp sign changes. This differential wrapping is a general feature of any perturbation, but in the case of large impacts, the amplitude of the velocity oscillations is substantial enough to make these sharp changes visually prominent. In Sect.\,\ref{sect:analytic_derivation}, we use this intuition to find particular PDE coefficients for the large impact case.
These features are also clear in 1D projections (Fig.\,\ref{fig:velocity_maps_large_r}), showing triangular waves in $V_R$ and sawtooth shapes in $V_\phi$. Comparison with the upscaled linear limit (dashed blue line) highlights the non-linear effects discussed in Sect.\,\ref{sect:large_impact_analytic}. These 1D projections, which resemble observed features in \emph{Gaia} data \citepalias{antoja2022tidal}, are useful for comparison with observations.

%
%

\section{Dynamical Discovery with SINDy}\label{sect:SINDy}

Here, we apply SINDy to the simulation time-series data to \mbox{(re-)discover} their governing equations.

\subsection{SINDy}

\begin{figure*}
    \centering
    \includegraphics[width=0.94\linewidth]{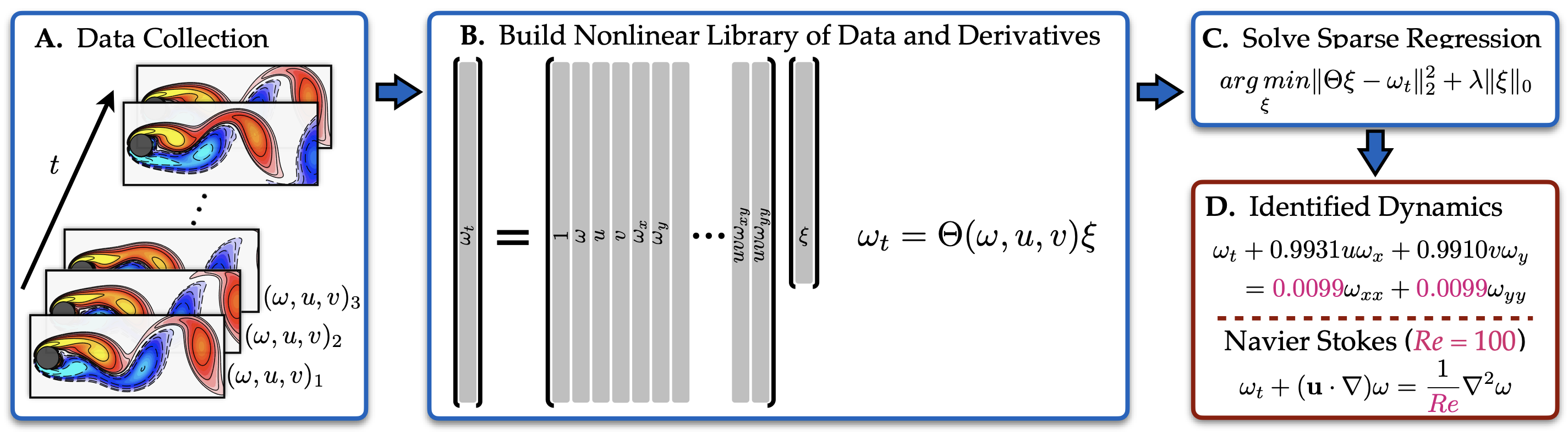}
    \caption{Schematic example of the usage of SINDy to infer the Navier-Stokes equation in a simulation. \textit{A}. Collect snapshots representing the solution of the PDE. \textit{B}. Compute numerical derivatives and organize the data into a comprehensive matrix $\Theta$ that includes the candidate terms for the PDE. \textit{C}. Employ sparse regression techniques to isolate the active terms in the PDE. \textit{D}. Active terms in the library ($\mathbf{x}(t)$, in the text) are informative of the underlying PDE. Reproduced from \citet{brunton2023mlpde}, who adapted it from  \citet{rudy2017sindypde}.}
    \label{fig:sindy_diagram}
\end{figure*}

SINDy \citep[][]{brunton2016sindy} is a data‐driven method that extracts the governing equations of a dynamical system directly from measured or simulated data. It is based on the idea that even highly non-linear systems can often be described by equations with only a few dominant terms.

For example, a standard linear ODE system is given by
\begin{equation}
    \frac{d \mathbf{x}}{dt}(t) = \mathbf{\Theta}\,\mathbf{x}(t),
\end{equation}
where $\mathbf{x}(t)$ is the state vector and $\mathbf{\Theta}$ is a constant matrix that characterizes the system dynamics. In practice, SINDy extends this framework to non-linear systems by constructing a library $\mathbf{x}(t)$ of non-linear candidate functions (e.g. polynomials, trigonometric functions, etc) and then uses sparse regression to identify the key terms that govern the behaviour. This approach yields a parsimonious model that preserves the physical interpretability of the system. Importantly, the output of SINDy is limited to identifying the best fit among terms provided in its library.

While originally designed for ODEs ($\dot{\mathbf{x}} = f(\mathbf{x})$), the core sparse regression idea extends to PDEs $u_t = N(u, u_x, u_{xx}, \dots)$ by utilizing spatial-temporal data \citep{rudy2017sindypde,schaeffer2017pde}. The process involves numerically computing the time derivative ($u_t$) and relevant spatial derivatives (e.g. $u_x, u_{xx}$) from the measured field $u(x,t)$ across all sampled locations and times. These computed derivatives, along with the non-linear combinations of the variables in $u$, form the library of candidate terms for the right-hand side of the PDE. A summary and example of the methodology is shown in the diagram of Fig.\,\ref{fig:sindy_diagram}, applied to the Navier-Stokes PDE.

\subsection{Equations for the small impact evolution}

We now apply SINDy to the small impact simulations (Sect. \ref{sect:m2_kick}, \ref{sect:generalized_kick}). The results of this section are independent of the shape of the initial kick, as long as it is small.

\begin{figure}
    \centering
    \includegraphics[width=0.94\linewidth]{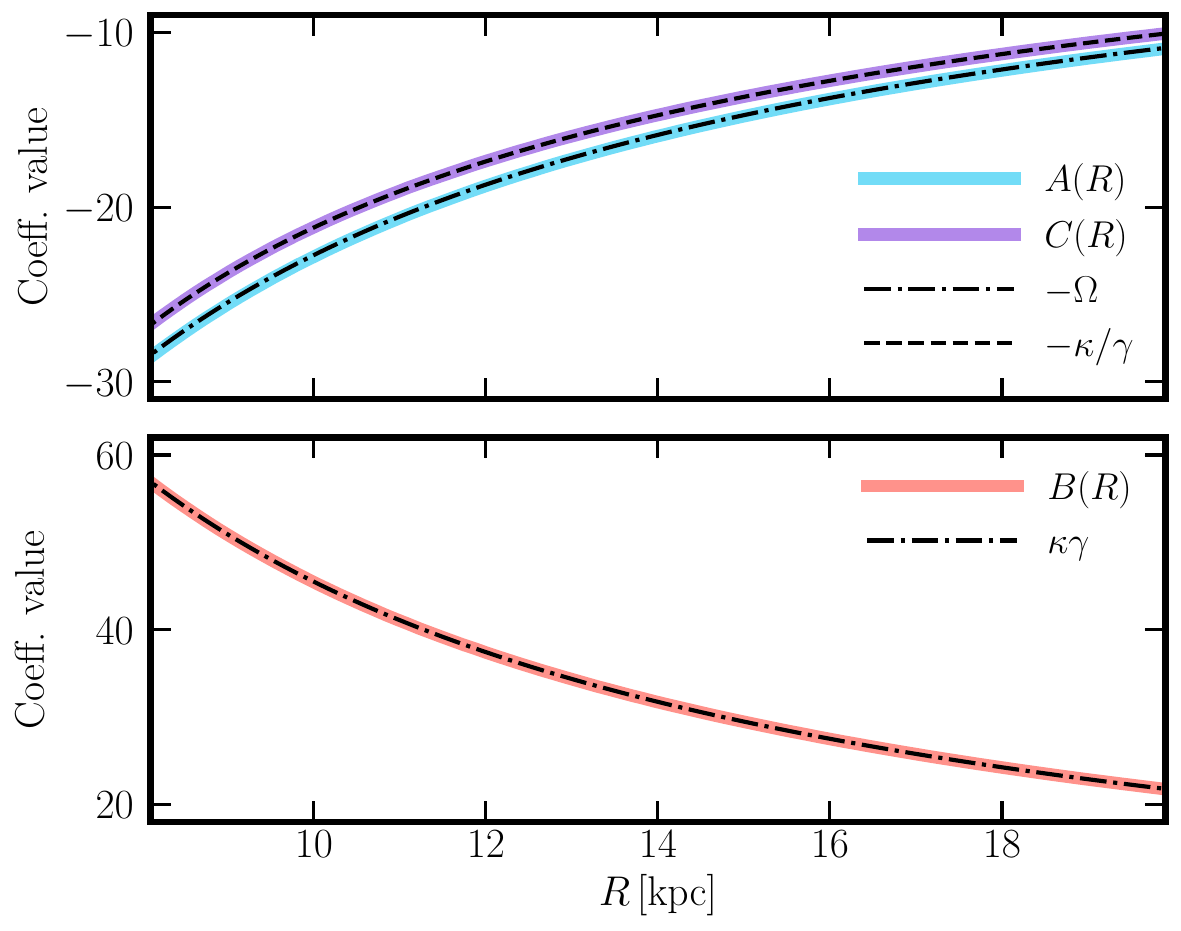}
    \caption{Radial profiles of the coefficients $A$ (blue), $B$ (coral), and $C$ (purple) from the linear PDE system (Eq.\,\ref{eq:lin_sys_sindy}) found by SINDy, governing the evolution of velocity perturbations in small impact simulations (Eq.\,\ref{eq:m_kick}). The black lines show the corresponding analytical values derived in Sect.\,\ref{sect:analytic_derivation} (Eq.\,\ref{eqn:lin_syst}). The excellent agreement between the SINDy-discovered coefficients and the analytical predictions demonstrates the success in recovering the underlying linear dynamics.}
    \label{fig:sindy_results}
\end{figure}

We treat the velocity perturbations, $\Delta V_R$ and $\Delta V_\phi$, as spatial and time dependent fields, $u(R, \phi, t) = (\Delta V_R, \Delta V_\phi)$. To apply the SINDy PDE approach, we compute the spatial derivatives ($\partial u / \partial R$, $\partial u / \partial \phi$) numerically from the simulation data\footnote{Spatial derivatives were computed numerically by fitting a polynomial of order $3$ in a window of $200$\,pc in the radial direction, and $6^\circ$ in the azimuthal direction. These values offer a robust combination of low noise and fast computation, but in any case the results do not depend on these parameters.}. Our library is composed of polynomial combinations, up to second order, of the state variables and their spatial derivatives:
\begin{equation}
    \boldsymbol{x}(t) = \bigg(\Delta V_R, \Delta V_\phi, \frac{\partial V_R}{\partial R}, \Delta V_R \Delta V_\phi, (\Delta V_\phi)^2,...\bigg),
\end{equation}
resulting in a library $\mathbf{x}(t)$ of $36$ terms. SINDy is then used to identify the sparse combination of candidate terms that best approximates the measured time derivatives across the entire dataset. Since the underlying gravitational potential is axisymmetric, we expect the coefficients of the governing equations to be independent of $\phi$, and depend only on $R$. To capture this, we run SINDy separately at each radial bin ($R$). For each radial bin, we combine the data from all $\phi$ locations within that bin (across all times) to build the SINDy regression problem, allowing us to determine coefficients specific to that radius. 

SINDy consistently identified that the dynamics are best described by a linear coupling between the velocity components and their $\phi$-derivatives. Specifically, the approach yielded the following equations:
\begin{align}\label{eq:lin_sys_sindy}
    \frac{\partial \Delta V_R}{\partial t} &= A(R) \frac{\partial \Delta V_R}{\partial \phi} + B(R) \Delta V_\phi, \\
    \frac{\partial \Delta V_\phi}{\partial t} &= A(R) \frac{\partial \Delta V_\phi}{\partial \phi} + C(R) \Delta V_R, \nonumber
\end{align}
where $A(R)$, $B(R)$, and $C(R)$ are numerical coefficient curves extracted by SINDy at each radial bin. Fig.\,\ref{fig:sindy_results} shows the numerical profiles of these curves (along with a spoiler of the theoretical values we find in Sect.\,\ref{sect:analytic_derivation}). We tested the approach with a variety of potentials and ICs, and observed that the extracted $A(R)$, $B(R)$, and $C(R)$ curves are robust against the different $m$ modes, but do depend on the underlying potential shape. The match between the left side (temporal derivatives) and right side (SINDy library) of both lines in Eq.\,\ref{eq:lin_sys_sindy} is excellent ($r^2\,\sim\,1$), where the squared correlation coefficient $r^2$ measures the linear correlation between the numerical temporal derivatives and the prediction obtained using the SINDy coefficients, thus showing that SINDy perfectly recovers the governing equations of this system, which motivates our analytical interpretation in Sect.\,\ref{sect:analytic_derivation}.

\subsection{Equations for the large impact}\label{sect:sindy_large}

\begin{figure}
    \centering
    \includegraphics[width=0.98\linewidth]{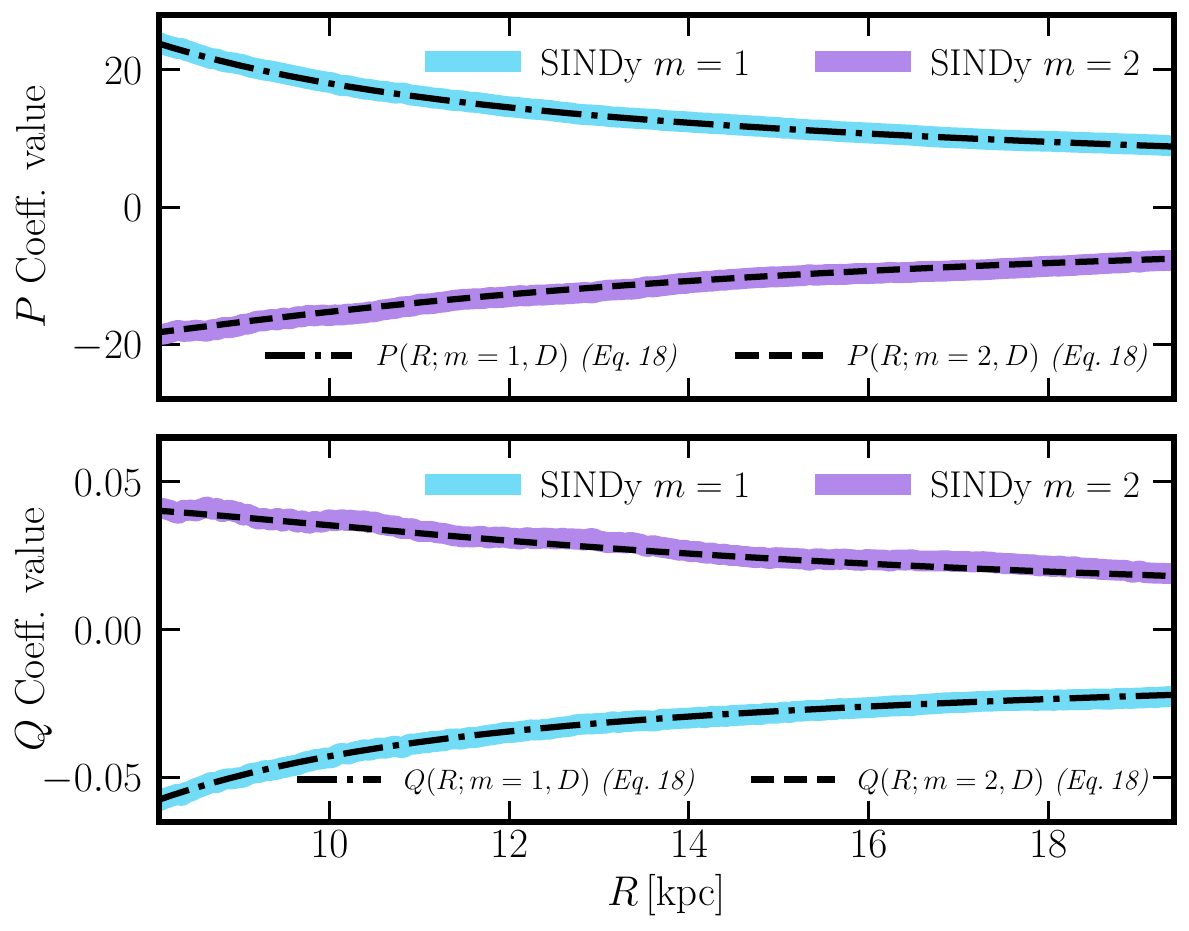}
    \caption{$P(R;m,D)$ and $Q(R;m,D)$ coefficients found by SINDy for the large impact (Eq.\,\ref{eq:nonlin_sys_sindy}) and the corresponding analytical value (Sect.\,\ref{sect:large_impact_analytic}, Eq.\,\ref{eq:large_impact_coefs}). Since the solution depends on $m$ and $D$, here we show examples for $D = 0.4$ and $m=1$ (purple) and $m=2$ (blue).}
    \label{fig:sindy_large}
\end{figure}

We now perform the SINDy analysis for the large impact simulations (Sect.\,\ref{sect:large_impact_sims}, Fig.\,\ref{fig:velocity_maps_large_xy}). In this case, where small deviations close to equilibrium can no longer be assumed, we apply SINDy directly to the velocity fields, $V_R$ and $V_\phi$. The library $\mathbf{x}(t)$ is thus constructed using $V_R$, $V_\phi$, their spatial derivatives ($\partial/\partial R$ and $\partial/\partial \phi$), and all their second-order polynomial combinations, resulting also in $36$ terms.

SINDy extracts a system of equations with a different combination of parameters:
\begin{align}\label{eq:nonlin_sys_sindy}
    \frac{\partial V_R}{\partial t} &= P(R;m,D) \frac{\partial V_R}{\partial \phi} + Q(R;m,D)  V_\phi \frac{\partial V_R}{\partial \phi}, \\
    \frac{\partial V_\phi}{\partial t} &= P(R;m,D) \frac{\partial V_\phi}{\partial \phi} + Q(R;m,D)  V_\phi \frac{\partial V_\phi}{\partial \phi}, \nonumber
\end{align}
where $P(R;m,D)$ and $Q(R;m,D)$ are numerical coefficients, shown in Fig.\,\ref{fig:sindy_large}. The excellent fit ($r^2 \sim 1$) indicates that the model captures the dynamics of the system. These coefficients depend not only on the radius and potential but also on the characteristics of the initial impact, namely its shape (mode $m$) and strength ($D$), as shown in Fig.\,\ref{fig:sindy_large}, where $m=1$ (blue) coefficients are different than the $m=2$ ones (purple). Consequently, the coefficients in Eq.\,\ref{eq:nonlin_sys_sindy} are specific to the dynamics established by the initial impact parameters ($m, D$). The method thus captures the dynamics of a specific impact, not a universal law. In Sect.\,\ref{sect:large_impact_analytic}, we exploit the simple quasilinear structure of Eq.\,\ref{eq:nonlin_sys_sindy} to describe it analytically.

%
%

\section{Interpreting tidal spirals through the learned equations}\label{sect:analytic_derivation}

In the previous section, we found equations describing the evolution of tidal spirals resulting from small (Eq.\,\ref{eq:lin_sys_sindy}) and large impacts (Eq.\,\ref{eq:nonlin_sys_sindy}). SINDy functions provide candidate mechanisms that can then be further analysed and justified through additional experimentation and reasoning. In this way, SINDy serves as a tool to guide us in developing enhanced models and identifying relevant physical mechanisms.

\subsection{Small impact equations}\label{sect:small_impact_analytic}

For small impacts, the linear PDE system (Eq.\,\ref{eq:lin_sys_sindy}) can be derived from linearising the Jeans equations, derivable from the Collisionless Boltzmann Equation (CBE). A complete derivation of this system is given in Appendix\,\ref{app:analytical_derivations_all}. We reproduce here the resulting linear PDE system for convenience:
\begin{eqnarray}\label{eqn:lin_syst}
\frac{\partial \Delta V_R}{\partial t} &=& -\Omega \frac{\partial \Delta V_R}{\partial \phi} + \kappa \gamma \Delta V_\phi,  \\
\frac{\partial \Delta V_\phi}{\partial t} &=& -\Omega \frac{\partial \Delta V_\phi}{\partial \phi} - \frac{\kappa}{\gamma} \Delta V_R. \nonumber
\end{eqnarray}
These reveal the analytic explanation of the coefficient curves obtained by SINDy in Eq.\,\ref{eq:lin_sys_sindy}:
\begin{equation}
    A(R) = -\Omega,\quad B(R) = \kappa\gamma, \quad C(R) = -\frac{\kappa}{\gamma}.
\end{equation}
In Fig.\,\ref{fig:sindy_results}, we show these analytical values on top of the SINDy results.

Equation \ref{eqn:lin_syst} is equivalent to Eqs. 2b and 2c in \citet{linshu1964} and Eqs. 10 and 11 in \citet{toomre1964linearized}. While similar to these classic equations, our approach differs: they studied self-gravitating propagation from equilibrium ICs, whereas we neglect self-gravity and impose an initial kinematic kick. The analytical solution to this system (see Appendix \ref{app:analytical_solution}) is a sum of spiral kinematic waves with frequencies $\omega = m\Omega \pm \kappa$:
\begin{eqnarray}\label{eq:fourier_solution}
\Delta V_R(R, \phi, t) = \sum_m \Big[ a_m e^{i(m\phi - (m\Omega + \kappa)t)} + b_m e^{i(m\phi - (m\Omega - \kappa)t)} \Big],\\
\Delta V_\phi(R, \phi, t) = \sum_m \Big[ c_m e^{i(m\phi - (m\Omega + \kappa)t)} + d_m e^{i(m\phi - (m\Omega - \kappa)t)} \Big],\nonumber
\end{eqnarray}
where $a_m, b_m, c_m, d_m$ are amplitudes determined by the initial conditions, and $c_m = -\frac{i}{\gamma} a_m$, $d_m = \frac{i}{\gamma} b_m$.
These waves wrap at pattern speeds
\begin{equation}
    \Omega_p = \Omega \pm \frac{\kappa}{m},
\end{equation}
demonstrating the relation (Eq.\,\ref{eq:empirical_omk}) that we discussed in Section\,\ref{sect:generalized_kick}. This solution provides an explicit, closed-form description of the temporal evolution of the residual velocities, setting the stage for interpreting the underlying dynamics.

For the weak, impulsive $m=2$ case \citep{struck2011tidal} discussed in Sect.\,\ref{sect:m2_kick}, applying the ICs (Eq.\,\ref{eq:m2_kick}) to this framework (see Appendix \ref{app:m2_impact}) shows that the kick decomposes into two waves wrapping at $\Omega \pm \kappa/2$. Their relative amplitude is determined by the potential:
\begin{equation}\label{eq:amplitudes_analytical}
    \frac{\mathcal{A}_{\Omega - \kappa/2}}{\mathcal{A}_{\Omega + \kappa/2}} = \frac{|b_2|}{|a_2|} = \frac{1 + \gamma}{\gamma-1}.
\end{equation}
For our fiducial potential, with a $\gamma \approx 1.46$ across the disc, we obtain a ratio of $\sim 5.34$, which matches our empirical measurement in Sect.\,\ref{sect:generalized_kick}. Fig.\,\ref{fig:small_1d} shows the 1D velocity profiles, illustrating the dominance of the $\Omega - \kappa/2$ mode.

\subsection{Large impact equations}\label{sect:large_impact_analytic}

Here we analyse the non-linear PDE (Eq.\,\ref{eq:nonlin_sys_sindy}) discovered by SINDy in the large impact simulation. Unlike in the small impact case where the SINDy-discovered linear system (Eq.\,\ref{eq:lin_sys_sindy}) directly corresponds to the linearized Jeans equations, the non-linear PDE (Eq.\,\ref{eq:nonlin_sys_sindy}) found for the large impact is, as discussed in Sect.\,\ref{sect:sindy_large}, specific to the ICs and not a general law. Thus, we do not expect a straightforward derivation from first principles. Here we focus on interpreting the dynamics directly through the structure of the PDE identified by SINDy for this particular type of strong perturbation.

We focus our analytical analysis on the azimuthal equation because its structure is ``self-contained'', meaning its form depends only on $V_\phi$ and its spatial derivatives, allowing us to determine the evolution of $V_\phi$ independently of $V_R$.
SINDy found that both equations in the system (Eq.\,\ref{eq:nonlin_sys_sindy}) share the same coefficients, $P$ and $Q$. This means that we can derive these key coefficients, applicable to both velocity components, by analysing only one of the equations.

The azimuthal part of Eq.\,\ref{eq:nonlin_sys_sindy} is a first-order quasilinear PDE solved by the method of characteristics. This means that the quantity $V_\phi$ remains constant along characteristic curves defined by 
\begin{equation}\label{eq:characteristic_curve}
    \frac{d\phi}{dt} = P(R;m,D) + Q(R;m,D) V_\phi, \quad \frac{d V_\phi}{dt} = 0.
\end{equation}
In Sect.\,\ref{sect:large_impact_sims} we explained that, at each $R$, the minimum $V_\phi$ corresponds to the apocentre of an inner orbit with guiding radius $R_{\text{g, in}}$, and moves at a constant pattern speed $\Omega(R_{\text{g, in}}) - \kappa(R_{\text{g, in}})/m$. Similarly, the points where $V_\phi = V_c$ move at a constant pattern speed $\Omega(R) - \kappa(R)/m$.
These associations provide two constraints on the coefficients $P$ and $Q$. By equating the instantaneous pattern speed from the characteristic curve (Eq.\,\ref{eq:characteristic_curve}) to the expected pattern speed for two specific constant 
$V_\phi$ values, we arrive at the following system of equations:
\begin{eqnarray}
P + Q\left(V_c - \frac{D}{\gamma} R\right) &=& \Omega(R_{\text{g,in}}) - \frac{\kappa(R_{\text{g,in}})}{m},\\
P + Q V_c &=& \Omega(R) - \frac{\kappa(R)}{m},\nonumber
\end{eqnarray}
where the first equation corresponds to the apocentres ($V_\phi = V_c - D R / \gamma$, minimums in Eq.\,\ref{eq:m_kick_large}), that evolve with pattern speed $\Omega(R_{\text{g,in}}) - \kappa(R_{\text{g,in}})/m$ and the second equation corresponds to points where $V_\phi = V_c$, with pattern speed $\Omega(R) - \kappa(R)/m$.

Solving this system yields a closed form for the PDE coefficients
\begin{eqnarray}\label{eq:large_impact_coefs}
    Q(R;m,D) &=& -\frac{\gamma}{D R} \left( \Delta \Omega - \frac{\Delta \kappa}{m} \right),\\
    P(R;m,D) &=& -\left( \Omega(R) - \frac{\kappa(R)}{m} \right) - Q V_c,\nonumber    
\end{eqnarray}
where we defined $\Delta \Omega \equiv \Omega(R_{\text{g, in}}) - \Omega(R)$ and $\Delta \kappa \equiv \kappa(R_{\text{g, in}}) - \kappa(R)$. In Fig.\,\ref{fig:sindy_large}, we show that these curves match perfectly with the coefficients extracted by SINDy, providing strong validation for both the SINDy discovery and our analytical interpretation.

\paragraph{Crossing time}

The characteristic curves of the system reveal an important dynamical aspect: their eventual crossing. In the astrophysical context, this corresponds to the appearance of overlapping kinematic populations in the same spatial volume, which challenges the definition of a unique ``mean velocity'' and, ultimately, the assumptions in the Jeans equations.

From \citet{zachmanoglou1986shock}, we can extract that, for a PDE with the shape of Eq.\,\ref{eq:nonlin_sys_sindy}, this overlap or crossing time is:
\begin{equation}
    t_{\text{cross}} = -\frac{1}{Q_{m,D}(R)\, V_{\phi,0}'(\phi_c)},
\end{equation}
meaning that the earliest crossing originates at the point $\phi_c$ where the derivative of the initial condition $V_{\phi,0}$ (Eq.\,\ref{eq:m_kick_large}) is the most negative. This gives us a closed expression for the crossing time 
\begin{equation}\label{eq:shock}
    t_{\text{cross}} = \frac{1}{m \Delta \Omega - \Delta \kappa}.
\end{equation}

Even in simulations where the Sgr impact on the velocities at the Solar neighbourhood is on the larger side of the spectrum ($\Delta V_R,\,\Delta V_\phi\,\sim\,25\,\kms$), this would translate to $D\,\sim\,2$. In our framework, this yields $t_{\rm cross}\gtrsim1.5$\,Gyr, comfortably exceeding the estimate of the last Sgr passage ($\lesssim 1\,$Gyr). Indeed, \citet{hunt2024rvr_spiral} showed that, for a kick compatible with a Sgr passage, the wraps of the radial phase spiral appear at least after $\sim1$\,Gyr. Thus, in the context of a Sgr passage, we do not expect a crossing of the perturbed populations.

%
%

\section{Applying SINDy to more realistic simulations}\label{sect:n-body}

\begin{figure*}
    \centering    
    \includegraphics[width=0.99\linewidth]{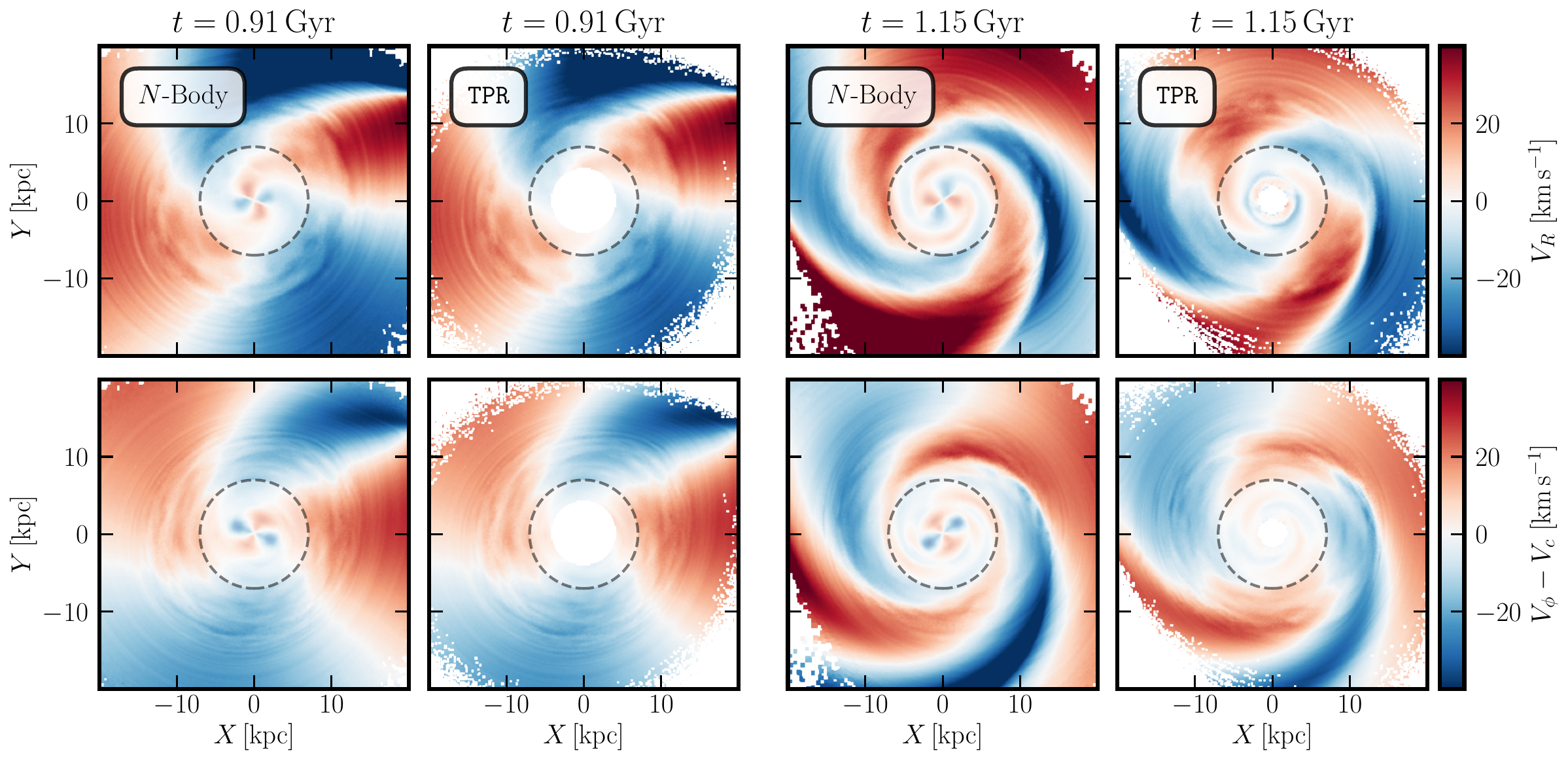}
    \caption{Kinematic maps comparing the evolution of velocity fields in the $N$-body (first and third columns) and \texttt{TPR} models (second and fourth columns). The panels show snapshots of the radial velocity $V_R$ (top row) and the azimuthal velocity perturbation $\Delta V_\phi$ (bottom row) in the $X$-$Y$ plane at two different times. The dashed circle denotes the inner region ($R < 7$\,kpc) that was excluded from the analysis. Observe the winding of the spiral patterns over time and the overall similarity between the $N$-body and \texttt{TPR} models, with subtle differences, particularly in the amplitude of the kinematic waves.}
    \label{fig:nbody_velocity_maps}
\end{figure*}

\begin{figure}
    \centering 
    \includegraphics[width=0.99\linewidth]{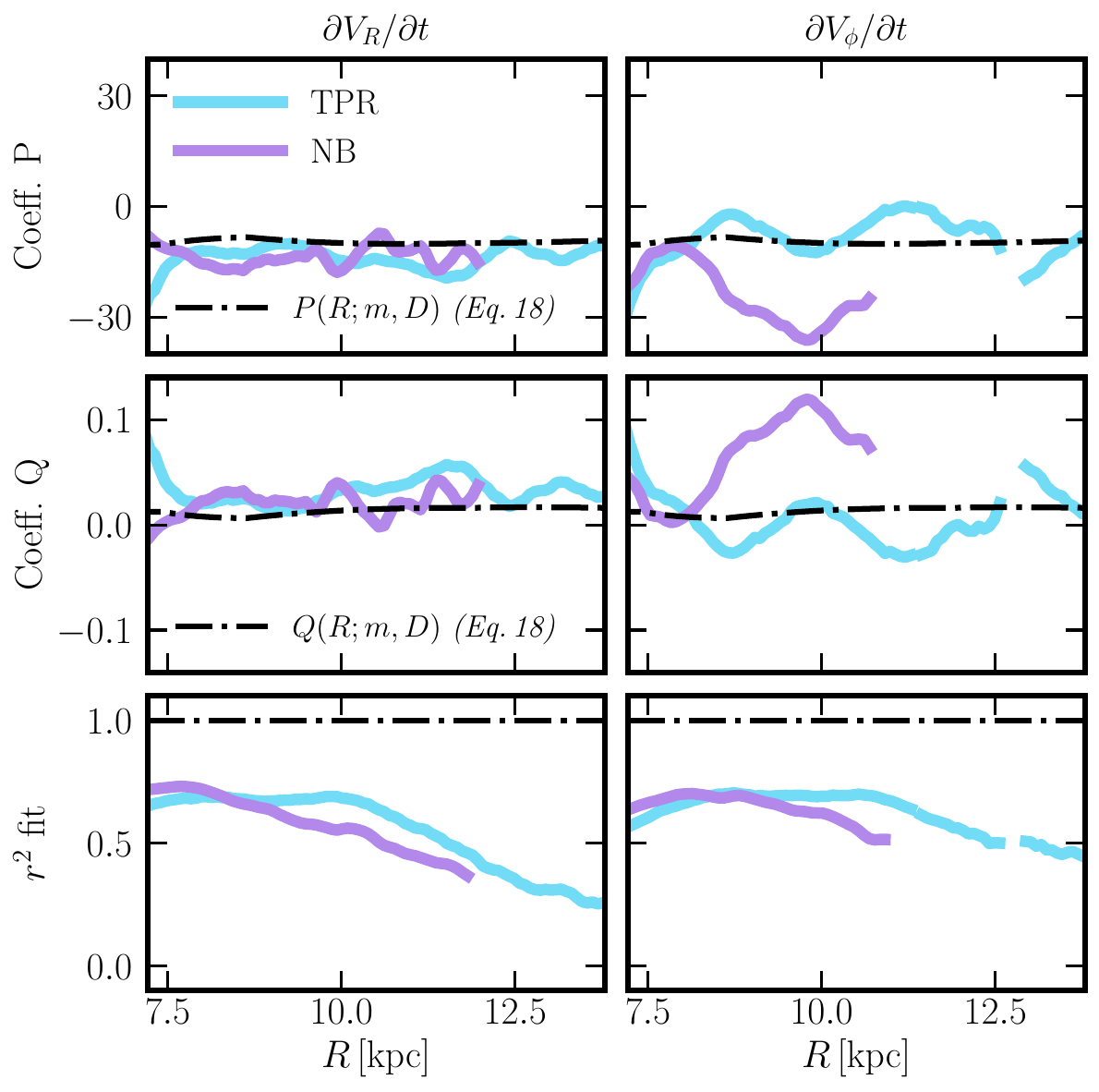}
    \caption{Results from applying SINDy to the \texttt{TPR} (blue lines) and $N$-body (purple lines) models. The top row shows the SINDy-recovered coefficient $P$, and the middle row shows coefficient $Q$, both as a function of $R$. The bottom row shows the $r^2$ value, indicating the goodness-of-fit for the SINDy model at each radius. The black dash-dotted lines represent the analytical predictions for $P$ and $Q$ (Eq.\,\ref{eq:large_impact_coefs}). The coefficients $P$ and $Q$ obtained by fitting the radial derivative ($\partial V_R / \partial t$) and the azimuthal derivative ($\partial V_\phi / \partial t$) are shown in the left and right columns, respectively. Voids in the curves correspond to radii where SINDy failed to find a solution.}
    \label{fig:sindy_nbody}
\end{figure}

We now test our method on a more realistic $N$-body simulation of a Sgr-like impact and a test-particle rerun (\texttt{TPR}) of the same simulation \citep[][Asano et al., in prep.]{asano2025sim_sgr}, initialized with particles from the same snapshot but evolved in a fixed axisymmetric potential. Running SINDy on these models allows us to test the method in more complex 3D setups with velocity dispersion and self-gravity. 
Details of these simulations can be found in Appendix\,\ref{app:realistic_sim_details}.

Figure \ref{fig:nbody_velocity_maps} shows the velocity maps for the $N$-body (first and third columns) and \texttt{TPR} (second and fourth columns) models over a $\sim 250\,$Myr period after the first Sgr pericentre. A bar dominates the inner disc ($R<7$\,kpc), so we focus on the outer region ($R>7$\,kpc). In this outer region, we see similar patterns to those discussed in Sect.\,\ref{sect:large_impact_sims} with our idealized test particle models with large impact. We observe a winding spiral with pattern speed $\Omega - \kappa /m$, which we measured using Fourier decomposition. This spiral pattern also displays characteristic non-linear features: regions with negative $V_R$ (blue) tend to concentrate, and the $V_\phi$ profile develops sharp edges. These realistic simulations, however, also reveal additional substructure (e.g., thin arches, bifurcations) that differs from the simplified idealized models. Finally, we observe subtle differences between these two models in Fig.\,\ref{fig:nbody_velocity_maps} , primarily a larger amplitude of the kinematic waves in the $N$-body simulation at $t=1.15$\,Gyr. This difference could be related to the self-gravity of the spiral arms and/or the extra acceleration of the perturber after the pericentre, which are not included in the \texttt{TPR} model.

\subsection{SINDy results}

We now apply the large-impact SINDy analysis (Sect. \ref{sect:sindy_large}) to the \texttt{TPR} and $N$-body models. In both models, from a library of $36$ terms, SINDy successfully recovered the exact same structural form of the non-linear PDE (Eq.\,\ref{eq:nonlin_sys_sindy}). 
This striking consistency shows the robustness of the method.

In Fig.\,\ref{fig:sindy_nbody} we show the $P$ (first row), and $Q$ (second row) coefficients obtained by SINDy in the \texttt{TPR} (blue) and $N$-body models (purple), as well as the analytical prediction (Eq.\,\ref{eq:nonlin_sys_sindy}).
The left (right) column shows the coefficients obtained when fitting the radial (azimuthal) part of the non-linear PDE (Eq.\,\ref{eq:nonlin_sys_sindy}). Although the coefficient $P$ should be identical in the radial and azimuthal equations (same for $Q$), SINDy fits them independently, allowing for different results. The bottom row of Fig.\,\ref{fig:sindy_nbody} shows the $r^2$ value. This value indicates how good the SINDy model fits the data at each radius. In both cases, the fit is sub-optimal ($r^2<0.7$). This points to missing terms in our library.

For the \texttt{TPR} model (blue lines in Fig.\,\ref{fig:sindy_nbody}), the recovered coefficients match the expected ones (dashed-dot lines, Eq.\,\ref{eq:large_impact_coefs}) fairly good, with the exception of some coherent oscillations in $R$.
For the $N$-body simulation (purple lines in Fig.\,\ref{fig:sindy_nbody}), the recovered coefficients are similar to the ones recovered from the TP rerun in the radial equation, but differ significantly in the azimuthal direction. We discuss the possible causes of this deviation in the following section.

\subsection{Limitations of the method}\label{sect:limitations}

The suboptimal $r^2$ fitting of the coefficients recovered by SINDy in the realistic models (bottom row in Fig.\,\ref{fig:sindy_nbody}) reveals significant limitations in our approach. Compared to the idealized setup (Sect. \ref{sect:large_impact_sims}), the realistic models introduce complexities such as vertical motion, velocity dispersion, different (more complex) initial conditions \citep[e.g.][]{semczuk2025tidal_tng}, and self-gravity. These factors are potentially responsible for the reduced goodness-of-fit.

\paragraph{Vertical motion}

Our analysis (Sect.\,\ref{sect:sindy_large}) uses a 2D model. However, the disc in the \texttt{TPR} and $N$-body models is 3D. We tested the impact of the vertical motion by applying SINDy to a subsample of particles with small $|Z|$ and $|V_Z|$ from the \texttt{TPR} model. The recovered coefficients were nearly identical to those from the full sample. This suggests that, in this specific simulation and time frame, the vertical motion is likely not the primary cause for the suboptimal fit of the equations.

\paragraph{Velocity dispersion}

In \citetalias{antoja2022tidal} we studied test particle simulations with a significant velocity dispersion, and showed that the pattern speed of the spiral arms and the evolution of the system is very similar to our idealised models. 
However, velocity dispersion fundamentally alters the dynamics by introducing pressure-like terms into the Jeans equations. This physical effect, which is absent in our cold-disc model, would require expanding the SINDy library to include terms related to the velocity dispersion tensor. The complexity is further compounded by the fact that real Galactic discs are composed of multiple stellar populations, each with its own characteristic velocity dispersion. 
Therefore, these factors likely contribute to the residual error and the suboptimal $r^2$ values observed in Fig.\,\ref{fig:sindy_nbody}.

\paragraph{ICs}

The ICs in the realistic simulations are taken at $t=0.9\,$Gyr, reflecting the full dynamical history (tidal impact, bar, previous spiral arms, etc.). This differs from the idealized analytic kick assumed for the theoretical coefficients (Eq.\,\ref{eq:m_kick_large}) that determines the values of the $P_{m,D}$ and $Q_{m,D}$ coefficients. Consequently, the recovered coefficients in Fig.\,\ref{fig:sindy_nbody} are not expected to perfectly match the analytical predictions across all radii.

\paragraph{Self-gravity}

The main discrepancy between the coefficients extracted from the \texttt{TPR} and $N$-body models is observed in the azimuthal equation (right column in Fig.\,\ref{fig:sindy_nbody}). We hypothesize that this discrepancy is driven by the self-gravity of the spiral arms. For loosely wound spirals, as in the temporal range we consider (Fig.\,\ref{fig:nbody_velocity_maps}), the self-gravity acceleration is more dominant in the azimuthal direction than in the radial direction \citep{linshu1964}. It is also important to notice that the radius where the difference between the \texttt{TPR} and $N$-body coefficients is larger, $R=10\,$kpc, corresponds to the Outer Lindblad Resonance radius of the bar. While this alignment may be coincidental, it raises the possibility that the bar is contributing to the discrepancy between the SINDy results and theoretical predictions.

\paragraph{Missing terms in the library}

The consistently low $r^2$ values in Fig.\,\ref{fig:sindy_nbody} (bottom row) are the strongest indication that the equation discovered by SINDy (Eq.\,\ref{eq:nonlin_sys_sindy}) is missing important physics, such as velocity dispersion and self-gravity.
While it is difficult to determine the precise contribution of each, these terms are necessary components of the full Jeans equations that govern disc dynamics, and capturing a more complete dynamical description in the future will require expanding the SINDy library to include them.

%
%

\section{The \emph{Gaia} $L_Z-$\mVR{ }wave}\label{sect:gaia_wave}

\begin{figure*}
    \centering 
    \includegraphics[width=0.99\linewidth]{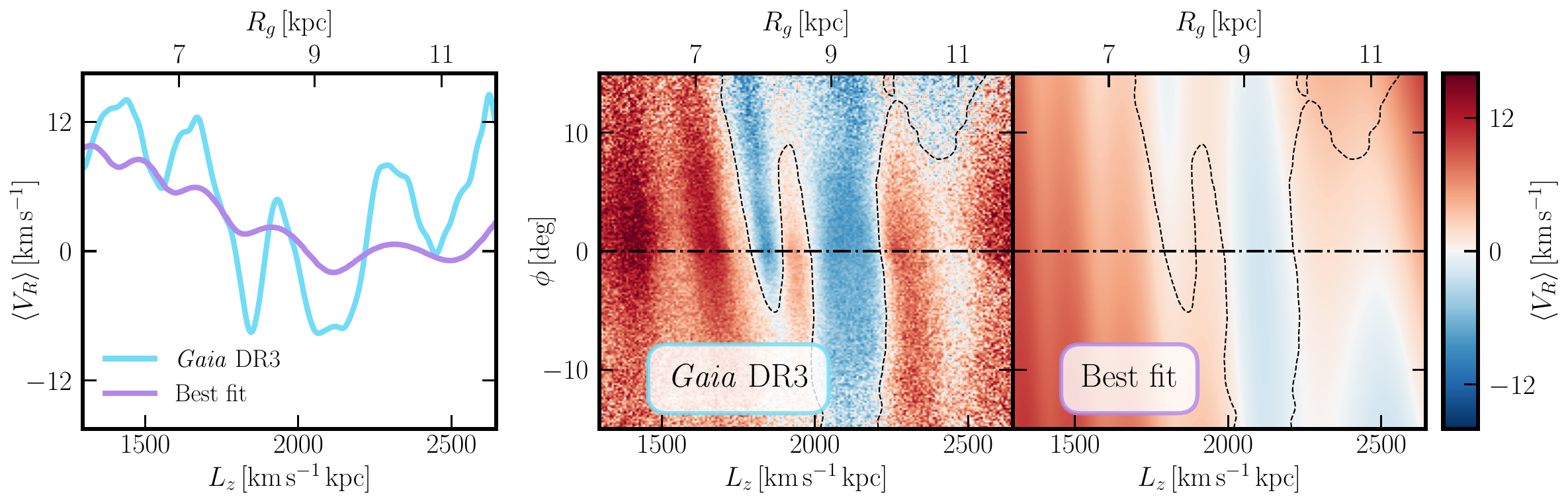}
    \caption{Best fit of an impulsive $m=2$ perturbation to the \emph{Gaia} DR3 $L_z-\phi-\langle V_R \rangle$ map. Left panel: \mVR{ }as a function of $L_z$ for \emph{Gaia} DR3 data (blue) and the best fit model (purple), at $\phi = 0^\circ$ (dot-dashed line in the other panels). Middle panel: The $L_z-\phi$ map of \mVR{ }for \emph{Gaia} DR3 data. Right panel: $L_z-\phi$ map of \mVR{ }for the best fit model. The dashed curve indicates the $\langle V_R \rangle = 0$\,km\,s$^{-1}$ curve of the data.}
    \label{fig:lz_vr_wave}
\end{figure*}

\citet{friske2019wave} showed a clear wave in the $L_z-V_R$ projection of the \emph{Gaia} data \citepalias[studied also in][\citealt{cao2024vrwave}]{antoja2022tidal}. They found a short and a long wavelength pattern with similar amplitudes ($\mathcal{A}_1 = 7.0\,\mathrm{km\,s^{-1}}$ and $\mathcal{A}_2 = 8.0\,\mathrm{km\,s^{-1}}$), and differing wavelengths  ($\lambda_1\,\sim\,285\,$kpc\,km\,s$^{-1}$ and $\lambda_2\,\sim\,1350\,$kpc\,km\,s$^{-1}$). The presence of these two waves, combined with the two wave pattern ($\Omega\pm\kappa/2$) observed in Sect.\,\ref{sect:simulations} from a single impact, raises a clear question: could we explain the full $L_Z-$\mVR{ }wave with a simple impulsive impact like that of a single Sgr passage?

To investigate this, we start by examining the $L_Z-$\mVR{ }wave in \emph{Gaia} DR3 data\footnote{We use StarHorse \citep{anders2022starhorse} distances, and typical quality cuts and Solar parameters. For more details, we refer to Sect.\,2.1 in \citet{bernet2024movinggroups}.} (blue curve in the left panel of Fig.\,\ref{fig:lz_vr_wave}). We also plot the mean radial velocity \mVR{ }in the $L_z-\phi$ plane \citep[central panel in Fig.\,\ref{fig:lz_vr_wave}, also shown in][\citetalias{antoja2022tidal}]{chiba2021bar}. We observe patterns consistent with previous findings, showing diagonal features in the $L_z-\phi$ plane, potentially indicative of spiral structure. We test the simplest scenario that could produce these signatures: a single distant impulsive Sgr passage (Eq.\,\ref{eq:m2_kick}). Using the analytical solution for an $m=2$ impulsive kick (derived in Appendix \ref{app:m2_impact}), we constructed $V_R = f(L_Z, \phi; t, \epsilon, \phi_0, V_{R,0})$ to fit the \mVR{ }of the observations. This model is a specific case of Eq.\,\ref{eq:fourier_solution} where the amplitudes of the two $m=2$ waves depend only on the impact strength $\epsilon$ (Eq.\,\ref{eq:params_simple}). Here, $t$ is the time since the impact, $\epsilon$ is the impact strength, $\phi_0$ is the initial angle of the perturber, and $V_{R,0}$ is a constant offset in $V_R$ \citep[equivalent to the $t$ parameter in Eq.\,2 in][]{friske2019wave}. The fit is done by minimizing the squared difference between the model and data\footnote{To find the optimal values for $t, \epsilon, \phi_0, V_{R,0}$, we run a simple search process. We sample different starting values to help avoid finding only a local best fit, and minimize the squared difference of the histograms (central and right panels in Fig.\,\ref{fig:lz_vr_wave}) using the L-BFGS-B implementation in \texttt{scipy.optimize} \citep{virtanen2020scipy}. We used \texttt{MilkyWayPotential2022} \citep{price-whelan2017gala} as the potential for the analytical formulation}.

The right panel of Fig.\,\ref{fig:lz_vr_wave} shows the best-fit model map in the $L_Z-\phi$ plane, and the purple curve in the left panel shows the corresponding $L_z-$\mVR{ }profile at $\phi=0^\circ$, compared to the \emph{Gaia} data (blue curve). The best values we found for the model parameters are:
\begin{equation*}\label{eq:gaia_fit_params}
   t = 0.42\,\text{Gyr},\,\,\epsilon=0.61\,\frac{\text{km\,s}^{-1}}{\text{kpc}},\,\, \phi_0=78^\circ, \,\, V_{R,0} = 6.3\,\text{km\,s}^{-1}. 
\end{equation*}
The values obtained are quite reasonable. For instance, the fitted time $t$ is roughly consistent with estimates for the time of impact derived from fitting the vertical phase spiral \citep{antoja2018phasespiral,frankel2023phasespiral,antoja2023phasespiral,darragh-ford2023phasespiral}. The initial angle $\phi_0$ is not well constraint by the fit.
For the impact parameter $\epsilon$, which can be approximated by $\epsilon = 2 G M_p /(r_p^2 V_p)$ (Sect.\,\ref{sect:m2_kick}), we see that, using the values for the last pericentre from \citet{vasiliev2021tango}, $r_p\,\sim\,24.3$\,kpc and $v_p\,\sim\,281$\,km\,s$^{-1}$, our fitted value implies a Sgr mass estimate of $M_p\,\sim\,1.17 \times 10^{10}$\,M$_\odot$, which is one order of magnitude larger than the total mass estimated at the last pericentre \citep{vasiliev2021tango}. The need for a higher mass of Sgr to reproduce the observables has already been pointed out in other studies \citep[e.g.][]{binney2018phasespiral,laporte2019phasespiral}. However, these mass estimates remain inconclusive due to the unconstrained mass loss history of Sgr.

More importantly, this model, and thus the obtained parameters, has several limitations. The impulsive kick model predicts a large amplitude difference between the two waves ($\sim\,5.34$,  Eq.\,\ref{eq:amplitudes_analytical}), while in the \emph{Gaia} data, these amplitudes are much more similar ($\sim\,1.14$). This means that the model cannot reproduce the observed balance between the two wave components. This is seen in the difference between the `short' waves in the left panel of Fig.\,\ref{fig:lz_vr_wave}. Another limitation is the use of the linear model instead of the non-linear equations. However, for the time and amplitude range of the fitting, we only expect small second-order corrections to the shape of the wave (see central panels of Fig.\,\ref{fig:velocity_maps_large_r}), and thus this would not be a significant issue. Finally, the limitations discussed in Sect.\,\ref{sect:limitations} also apply here, so accounting for the velocity dispersion or self-gravity effects could influence the fitting results.

In addition, our model is only considering a single, impulsive passage, but could be extended or combined with complementary dynamical processes. These could include the effect of multiple Sgr passages, each one generating one of the observed waves, as we explored in \citetalias[][]{antoja2022tidal}: closer non-impulsive perturbations; the combined influence of the Galactic bar and stellar spiral arms \citep[e.g.][]{hunt2019spiralphasemixing,khalil2024spiral_arm_pot}; or the cumulative effect of other past or present perturbers. Despite all the listed limitations, both the obtained fitting and parameters are promising results, pointing to a potential diagnostic of the Sgr passage using planar motions. We aim to explore this model in detail in future work.

%
%

\section{Discussion}\label{sect:discussion}

\subsection{Potential of the method}

Applying data-driven discovery tools like SINDy to simulations offers significant potential, serving as an empirical guide or providing simplified descriptions in complex regimes.
In the case of the small-impact simulations, SINDy identified the linear PDE system. This system is known to describe the evolution of perturbations in a collisionless disk \citep[e.g.][]{toomre1964linearized,linshu1964}. The significance here is not the novelty of the equation itself, but the ability of the method to extract this fundamental structure directly from the simulated phase-space data. This success served as a validation of the SINDy approach for this type of problem and provided the empirical foundation that guided our analytical derivation (Sect. \ref{sect:analytic_derivation}), enabling us to obtain explicit, closed-form solutions describing the phase-space distribution at all times (Eq.\,\ref{eq:fourier_solution}). 
The performance of SINDy depends critically on the temporal resolution of the data. While, obviously, the sampling must resolve the characteristic timescales of the system for SINDy to reproduce them, usually the challenge is ensuring the numerical stability of the time derivatives. In our analysis, we carefully checked that the fine time steps used were sufficient to obtain smooth and stable derivatives for both $V_R$ and $V_\phi$. In general, however, each application of the method will have its own requirements that need to be assessed.

For the specific large-impact scenario, SINDy discovered a non-linear PDE system (Eq.\,\ref{eq:nonlin_sys_sindy}) governing the kinematic perturbations. To our knowledge, this particular form has not been explicitly discussed previously in the context of galactic dynamics. Although this equation is specific to the used ICs and not a general law like the Jeans equations, its surprising simplicity – notably the absence of explicit radial derivatives and a single non-linear term – contrasts with the complexity of the full Jeans equations required to model this regime. This simpler structure directly facilitated the theoretical analysis of features like the characteristic curves and the derivation of the crossing time (Eq.\,\ref{eq:shock}). Applying SINDy to realistic simulations (Sect.\,\ref{sect:n-body}) recovered the same PDE structure, demonstrating the method robustness to velocity dispersion, 3D motion, and complex histories. The discrepancies in the recovered coefficients (Fig.\,\ref{fig:sindy_nbody}) are a valuable outcome, pointing to missing physics (like self-gravity, as discussed in Sect.\,\ref{sect:role_of_sg}) and guiding the systematic improvement of our models.

The governing PDEs we have identified also open up a route for efficient forward modelling. Instead of running full particle simulations or numerically solving PDEs directly, one can transform the PDEs into systems of coupled ODEs by expanding the dynamical fields in a basis of functions. This approach, characteristic of spectral methods, allows for faster evolution of the system in coefficient space, and is closely related to the latest developments in galactic dynamics \citep{rozier2019matrix,petersen2024linearresponse}. We demonstrate this technique for the linear PDE system and show its effectiveness in Appendix\,\ref{app:basis_function}. While more complex for the non-linear case, solving the resulting ODE system still offers significant computational advantages over traditional methods.

\subsection{Adding self-gravity to the mix}\label{sect:role_of_sg}

Our current simplified PDE (Eq.\,\ref{eq:nonlin_sys_sindy}) does not explicitly include terms representing the self-gravitational forces or the potential response of the live halo \citep{bernet2025darkspirals}. This introduced discrepancies when applying SINDy to more realistic simulations (Sect.\,\ref{sect:n-body}, Fig.\,\ref{fig:sindy_nbody}). The data-driven discovery framework, however, offers clear paths to address these limitations. One approach is to explicitly calculate the gravitational acceleration due to the evolving self-consistent density structure of the spiral arms (and potentially other components like the live halo or external perturber). These calculated accelerations could then be included as additional terms in the SINDy library, treating the self-gravity as an external force acting on the system.

Another clear path to extend the SINDy library is to include terms related to the density field itself. The method could potentially uncover a different, more comprehensive set of PDEs. We hypothesize that SINDy might identify a PDE structure that intrinsically accounts for the self-gravitational forces, potentially finding a more effective or simplified representation than the full, complex Jeans equations required for the coupled density-velocity system. 
Beyond self-gravity, these data-driven techniques could also prove powerful for uncovering tractable, scenario-specific models of complex effects such as dynamical friction—especially difficult to solve analytically—just as SINDy identified a relatively simple non-linear PDE (Eq.\,\ref{eq:nonlin_sys_sindy}) for the large-impact case. In addition, SINDy is closely related to other data-driven dynamical analysis methods, such as Dynamic Mode Decomposition \citep[DMD,][see \citealt{brunton2016sindy}]{schmid2010dmd}. DMD identifies dominant system modes and has been applied to simulations of vertical phase spirals \citep{darling2019dmd}.
In a similar direction, our methodology could be extended to model the full 3D response of the disc. Running the methodology in a similar setup, with a vertical (instead or in addition to the planar) perturbation, we could study the phase-mixing of bending waves, and their coupling with the planar dynamics, complementing, for example, the approaches of \citet{chequers2018subhalos,blandhawthorn2021phasespiral,asano2025sim_sgr}.

\subsection{Application to observational data}

In Section\,\ref{sect:gaia_wave}, we applied our analytical model for an impulsive kinematic kick (derived from the linear PDE, Eq.\,\ref{eq:fourier_solution}) to fit the observed $L_Z-$\mVR{ }wave in \emph{Gaia} data. Fig.\,\ref{fig:lz_vr_wave} shows that this simple model can reproduce the overall structure and winding pattern of the observed wave reasonably well, providing specific parameters for the Sgr passage time and strength, although several key limitations are mentioned. The idea of analysing kinematic wave structures like the $L_z - V_R$ wave can be generalised to model similar phenomena in external galaxies. The increasingly precise kinematic maps, obtained through integral field spectroscopy \citep[e.g.][]{erroz-ferrer2015ifu_noncircular,cappellari2016ifu_review,denBrok2020ifu_kinematics}, and even using \emph{Gaia} data in the LMC \citep{jimenez-arranz2023lmc_kin,jimenez-arranz2024lmc_bar,jimenez-arranz2025lmc_warp}, provide rich datasets of kinematic information. These datasets contain several examples of tidal features and kinematic waves, but cleanly distinguishing tidally-induced spirals from those generated by other mechanisms (e.g. bars or internal instabilities) remains a key open question.

The morphology of spiral arms is known to be linked to the properties of the host galaxy, exemplified by the correlation between pitch angle and galactic shear rate ($\Gamma$) in the context of transient, corotating spiral arms \citep[e.g.][]{linshu1964,seigar2005pitch,grand2013pitch}. The dynamics of the tidally-induced arms in our model are naturally consistent with this observational trend. The connection is direct: the pattern speeds we study, $\Omega_p = \Omega \pm \kappa/m$, are a function of shear via the relation $\kappa^2 \propto \Omega^2 (2 - \Gamma)$. Consequently, a higher shear rate leads to more rapid differential winding and intrinsically tighter arms\footnote{To build intuition, consider two extreme cases for an $m=2$ wave. In a solid-body disc ($\Gamma=0$, $\kappa=2\Omega$), the slow-wave pattern speed is $\Omega_p = \Omega - \kappa/2 = 0$, so the pattern is stationary and does not wind. In a Keplerian disc ($\Gamma=1.5$, $\kappa=\Omega$), the pattern speed is $\Omega_p = \Omega/2$.}. However, since the pitch angle in our model also depends on the time elapsed since the impact, we expect these arms to populate the relation with a significant scatter, corresponding to their evolutionary stage.

This evolutionary sequence implies a wide range of pitch angles, starting open and becoming progressively tighter. However, because the initial perturbation is purely kinematic, a density contrast only builds as the arms wrap, setting a practical maximum boundary on their observable pitch angle in density tracers. Conversely, the finite lifetime of the arm sets a minimum boundary. In our idealized model, this lifetime is marked by the kinematic winding into an unresolvable pattern or, for large impacts, by the non-linear crossing time ($t_{\rm{cross}}$, Eq.\,\ref{eq:shock}). In more realistic setups, this would be further accelerated by the velocity dispersion. Together, these effects define an observable window of pitch angles, though estimating its boundaries is complex, as it depends on the kick strength, host galaxy properties, and observational resolution. It is plausible that the observed spiral arm population is a mix of types, with the observed correlation possibly dominated by more isolated galaxies. This opens a powerful diagnostic avenue: using the morphology of tidal arms to independently constrain the gravitational potential of a galaxy and its interaction history.

%
%

\section{Summary and conclusions}\label{sect:conclusions}

The spiral structure and overall dynamics of the MW remain a major area of research, with recent observations highlighting that the MW disc is not in a steady state but is actively perturbed, likely by the passage of the Sgr dwarf galaxy. In this work, we explored the dynamics of tidally induced spiral arms applying SINDy, a data-driven technique that identifies the governing differential equations of a dynamical system directly from simulation data. Our main findings are summarized below:

\begin{itemize}
    \item A single $m$-fold impulsive velocity kick with small amplitude triggers the formation of two $m$-armed spiral waves, wrapping at pattern speeds $\Omega\pm\kappa/m$. The amplitudes of these waves depend on the shape of the initial kick and the potential, and can be analytically described (Sect.\,\ref{sect:analytic_derivation}).
    \item Applying SINDy to simulations of small impacts, we successfully recovered the linear PDE (Eq.\,\ref{eq:lin_sys_sindy}) governing the evolution of the system, validating the capability of the method in this context.
    \item Guided by the SINDy discovery, we analytically re-derived the linear PDE system from first principles using the CBE. This derivation recovers equations equivalent to classical descriptions \citep{toomre1964linearized,linshu1964} and provides a closed-form solution (Eq.\,\ref{eq:fourier_solution}) of the temporal evolution of the $V_R$ and $V_\phi$ maps.
    \item We defined a family of simplified larger impacts (Eq.\,\ref{eq:m_kick_large}) resulting in $m$-armed spiral patterns. For these cases, SINDy discovered a simple, previously unknown non-linear PDE system (Eq.\,\ref{eq:nonlin_sys_sindy}), that describes the evolution of the system, demonstrating its potential to uncover simplified governing equations in complex regimes.
    \item The coefficients of the non-linear PDE (Eq.\,\ref{eq:nonlin_sys_sindy}) are specific to the initial impact characteristics (mode $m$, strength $D$), highlighting that this equation describes the dynamics for a particular type of perturbation, not a universal law. However, its simple mathematical structure enables the interpretation of its characteristic kinematic features.
    \item The analysis of the non-linear PDE (Eq.\,\ref{eq:nonlin_sys_sindy}) provides a framework for interpreting the characteristic kinematic patterns observed in simulations (e.g. triangular $V_R$ and sawtooth $V_\phi$ profiles in Fig.\,\ref{fig:velocity_maps_large_r}), which resemble features in MW phase space (like the $R-V_\phi$ ridges and $L_z-V_R$ wave). Our interpretation of this PDE through characteristic curves reveals that these patterns arise from differential wrapping.
    \item Applying our data-driven framework to more realistic simulations (an $N$-body simulation and its test-particle rerun), we successfully recovered the same non-linear PDE structure (Eq.\,\ref{eq:nonlin_sys_sindy}) found in idealized simulations, demonstrating the potential of this method in more complex setups.
    \item The coefficients recovered from the realistic simulations showed discrepancies compared to idealized cases (Fig.\,\ref{fig:sindy_nbody}), which we attribute primarily to physical effects not explicitly included in our simplified PDE, such as self-gravity.
    \item We applied the closed-form solution (Eq.\,\ref{eq:fourier_solution}), assuming a simple impulsive kick, to fit the \emph{Gaia} $L_Z-$\mVR{ }wave (Sect.\,\ref{sect:gaia_wave}). We obtained reasonable visual fit and parameters for a Sgr passage, including a passage time of $\sim\,0.42$\,Gyr and an inferred perturber mass $\sim\,1.17 \times 10^{10}$\,M$_\odot$. However, our model shows a significant discrepancy in the amplitude ratio between the two main observed components. This mismatch suggests that the simple assumption of a single impulsive kick is an incomplete description of the $L_Z-$\mVR{ }wave. The discrepancy could arise from the limitations of this simplified model of the Sgr encounter, or it may indicate a more complex physical origin for the waves, such as multiple passages or the influence of other galactic components, like the bar.
\end{itemize}

The evolution of the field of galactic dynamics, increasingly data-driven, demands new methods to study the fundamental principles governing non-equilibrium dynamics.
Our approach offers a new path to explore theoretical dynamics often intractable otherwise. It complements other novel data-driven techniques in the field, such as Orbital Torus Imaging \citep[OTI,][]{price-whelan2021oti,price-whelan2025oti,horta2024oti,oeur2025oti} for constraining the gravitational potential, basis function expansions \citep{johnson2023bfe,arora2024forced,arora2025bfe} for efficient dynamical field evolution, and neural network-based potential recovery from kinematics \citep{green2023deep,kalda2024deep,zucker2025deep}. These frameworks that link empirical data to predictive analytical models offer a direct path for exploring theoretical dynamics and dissecting complex galactic structure and evolution, including challenging phenomena like self-gravity and dynamical friction.

\begin{acknowledgements}
We thank the anonymous referee for their constructive report.
We thank Kathryn V. Johnston, Chris Hamilton, Kiyan Tavangar, the EXP collaboration, especially the Bars \& Spirals working group, and the Gaia UB team for the valuable discussions that contributed to this paper.
We acknowledge the grants PID2021-125451NA-I00 and CNS2022-135232 funded by MICIU/AEI/10.13039/501100011033 and by ``ERDF A way of making Europe’’, by the ``European Union'' and by the ``European Union Next Generation EU/PRTR'', and the Institute of Cosmos Sciences University of Barcelona (ICCUB, Unidad de Excelencia ’Mar{\'\i}a de Maeztu’) through grant CEX2019-000918-M. 
MB acknowledges funding from the University of Barcelona’s official doctoral program for the development of a R+D+i project under the PREDOCS-UB grant. 
TA acknowledges the grant RYC2018-025968-I funded by MCIN/AEI/10.13039/501100011033 and by ``ESF Investing in your future''. 
This research in part used computational resources of Pegasus provided by Multidisciplinary Cooperative Research Program in Center for Computational Sciences, University of Tsukuba.
\end{acknowledgements}


\bibliographystyle{aa}
\bibliography{MyBib}


\appendix

\section{Radial profiles of the small impact}\label{app:radial_profiles}

\begin{figure*}
    \centering
    \includegraphics[width=.99\linewidth]{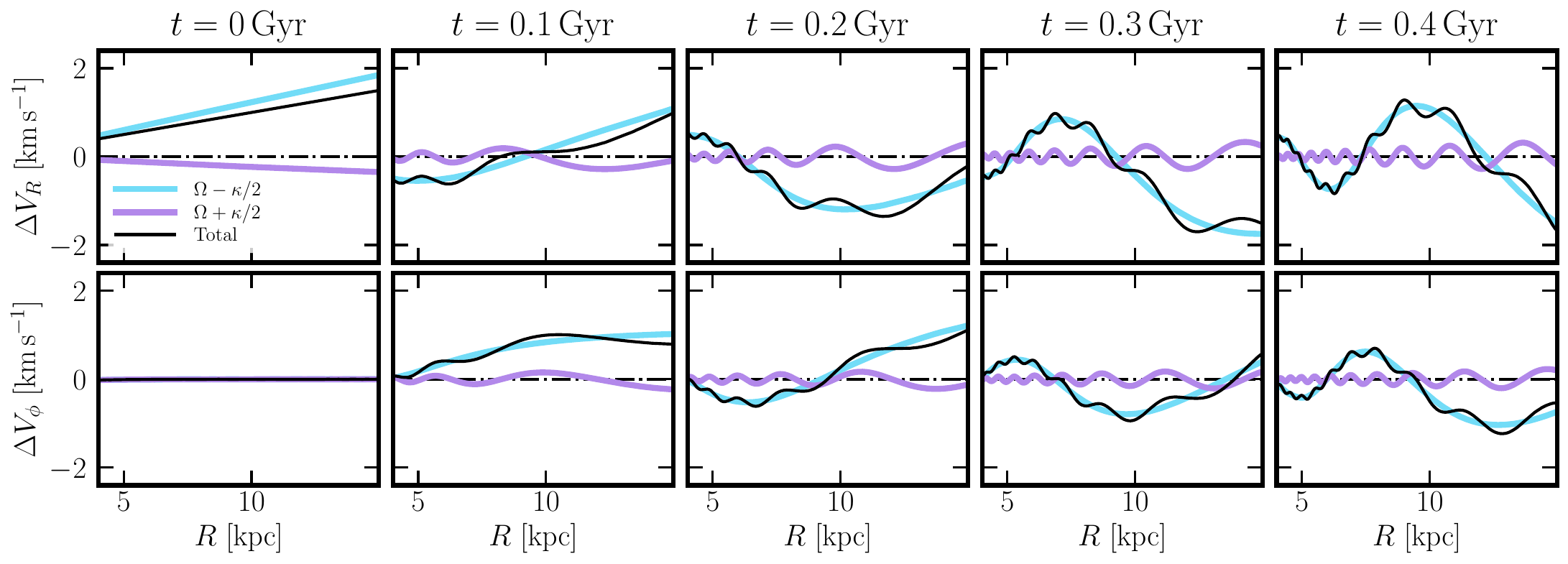}
    \caption{
       Radial profiles of velocity perturbations $\Delta V_R$ (top row) and $\Delta V_\phi$ (bottom row) for the small, distant, $m=2$ kinematic kick simulation (Sect.\,\ref{sect:m2_kick}), as a function of $R$, at four different times, at $\phi=0^\circ$. In each panel, the blue line represents the contribution from the $\Omega - \kappa/2$ wave, the purple line represents the $\Omega + \kappa/2$ wave, and the black line is the sum of these two modes, which corresponds to the total velocity perturbation at that radius. The initial kick ($t=0$\,Gyr) decomposes into the two modes. As time evolves, the $\Omega - \kappa/2$ mode remains dominant and shapes the overall perturbation pattern, while the $\Omega + \kappa/2$ mode appears as smaller-scale radial oscillations on top of it.
    }
    \label{fig:small_1d}
\end{figure*}

This appendix presents supplementary details and visualizations of the velocity structure resulting from the small, distant $m=2$ kinematic kick simulation discussed in Section \ref{sect:m2_kick}. Fig.\,\ref{fig:small_1d} shows the 1D radial profiles of the velocity perturbations $\Delta V_R$ (top row) and $\Delta V_\phi$ (bottom row) as a function of radius $R$, specifically along the line $\phi=0^\circ$, at five different times ($t=0, 0.1, 0.2, 0.3, 0.4$\,Gyr). Each panel breaks down the total velocity perturbation (black line) into the contributions from the two fundamental wave modes excited by the $m=2$ kick: the $\Omega - \kappa/2$ mode (blue line) and the $\Omega + \kappa/2$ mode (purple line). These profiles visually demonstrate how the initial kick at $t=0$\,Gyr is composed of these two modes and how the spatial structure evolves over time. Notably, the $\Omega - \kappa/2$ mode contributes significantly to the overall perturbation pattern, while the $\Omega + \kappa/2$ mode is typically seen as smaller-scale radial variations superimposed on the dominant mode.

\section{Analytical derivations}\label{app:analytical_derivations_all}

\subsection{Derivation of the linear equations}\label{app:analytical_derivations}

We start by considering the Collisionless Boltzmann Equation in 2D Cylindrical Coordinates, for an axisymmetric potential without self-gravity:
\begin{equation} \label{eq:CBE}
\frac{\partial f}{\partial t} + v_R \frac{\partial f}{\partial R} + \left( \frac{v_\phi^2}{R} - \frac{\partial \Phi}{\partial R} \right) \frac{\partial f}{\partial v_R} - \frac{v_R v_\phi}{R} \frac{\partial f}{\partial v_\phi} = 0
\end{equation}
Taking moments by multiplying by $v_R$ and $v_\phi$ and integrating over the velocities yields the Jeans Equation for the mean velocities $V_R$ and $V_\phi$. We assume there is no velocity dispersion, so all stress‐tensor terms drop out, matching the zero‑dispersion setup of our simulations (Sect.\,\ref{sect:simulations}). This yields:
\begin{eqnarray} \label{eq:jeans}
\frac{\partial V_R}{\partial t} + V_R \frac{\partial V_R}{\partial R} + \frac{V_\phi}{R} \frac{\partial V_R}{\partial \phi} - \frac{V_\phi^2}{R} + \frac{\partial \Phi}{\partial R} &=& 0, \\
\frac{\partial V_\phi}{\partial t} + V_R \frac{\partial V_\phi}{\partial R} + \frac{V_\phi}{R} \frac{\partial V_\phi}{\partial \phi} + \frac{V_R V_\phi}{R} &=& 0.\nonumber
\end{eqnarray}

We linearize the equation by assuming a small perturbation about the equilibrium of the form $V_\phi = V_c(R) + \Delta V_\phi$, $V_R = \Delta V_R$:
\begin{eqnarray}
\quad \frac{\partial \Delta V_R}{\partial t} + \Omega \frac{\partial \Delta V_R}{\partial \phi} - 2\Omega \Delta V_\phi &=& 0, \\
\quad \frac{\partial \Delta V_\phi}{\partial t} + \Omega \frac{\partial \Delta V_\phi}{\partial \phi} + \left( R \frac{d\Omega}{dR} + 2\Omega \right) \Delta V_R &=& 0,\nonumber
\end{eqnarray}
where $\Omega \equiv V_c/R = \sqrt{\frac{-1}{R}\frac{d\Phi}{dR}}$.
Assuming $\kappa^2 = R\frac{d\Omega^2}{dR} + 4\Omega^2$ and $\gamma = \frac{2\Omega}{\kappa}$, we can rewrite the linearized equations to obtain the following system of equations:
\begin{eqnarray}\label{eq:lin_syst_ap}
\frac{\partial \Delta V_R}{\partial t} &=& -\Omega \frac{\partial \Delta V_R}{\partial \phi} + \kappa \gamma \Delta V_\phi,  \\
\frac{\partial \Delta V_\phi}{\partial t} &=& -\Omega \frac{\partial \Delta V_\phi}{\partial \phi} - \frac{\kappa}{\gamma} \Delta V_R. \nonumber
\end{eqnarray}

In the following section, we derive the analytical solution of the linear PDE recovered here (Eq.\,\ref{eq:lin_syst_ap}).

\subsection{Solution of the linear equations}\label{app:analytical_solution}

For a fixed radius $R$, where $\Omega(R)$, $\kappa(R)$, and $\gamma(R)$ are constants, the linearized system (Eq.\,\ref{eq:lin_syst_ap}) becomes a pair of linear PDEs with constant coefficients. Exponential functions like $e^{i(m\phi - \omega t)}$ are eigenfunctions of linear differential operators, simplifying the PDEs to algebraic equations. Therefore, we can assume solutions of the form
\begin{equation}
    \Delta V_R = \Delta V_R^{m,\,\omega} e^{i(m\phi - \omega t)}, \quad \Delta V_\phi = \Delta V_\phi^{m,\,\omega} e^{i(m\phi - \omega t)},
\end{equation}
where $m$ is the azimuthal mode and $\omega$ is their frequency.
Substituting these solutions in the linearized system and operating it, we obtain
\begin{eqnarray}\label{eq:an_resolution_step}
(\omega - m\Omega) \, \Delta V_\phi^{m,\,\omega} &=& \frac{\kappa}{i\gamma} \, \Delta V_R^{m,\,\omega}, \\
(\omega - m\Omega) \, \Delta V_R^{m,\,\omega} &=& -\frac{\kappa \gamma}{i} \, \Delta V_\phi^{m,\,\omega},\nonumber
\end{eqnarray}
whose solution leads to the relation
\begin{equation}\label{eq:dispersion_relation}
    \omega = m\Omega \pm \kappa,
\end{equation}
which tell us which eigenfunctions ``survive'', i.e. which waves develop after the initial perturbation. The pattern speed of these waves ($\Omega_p \equiv \omega / m$) is then
\begin{equation}
    \Omega_p = \Omega \pm \frac{\kappa}{m}.
\end{equation}
Finally, solving Eq.\,\ref{eq:an_resolution_step} also gives the amplitude relation 
\begin{equation}\label{eq:eigenfunction_relation}
    \Delta V_\phi^{m,\,\omega} = \mp \frac{i}{\gamma} \Delta V_R^{m,\,\omega},
\end{equation}
where the inverted $\mp$ sign means that it is the opposite sign as the one obtained for $\Omega \pm \kappa/m$. 
Therefore, the analytical solution of the system consists of a sum of spiral density waves with frequencies $\omega = m\Omega \pm \kappa$:
\begin{eqnarray}\label{eq:fourier_solution_ap}
\Delta V_R(R, \phi, t) = \sum_m \Big[ a_m e^{i(m\phi - (m\Omega + \kappa)t)} + b_m e^{i(m\phi - (m\Omega - \kappa)t)} \Big],\\
\Delta V_\phi(R, \phi, t) = \sum_m \Big[ c_m e^{i(m\phi - (m\Omega + \kappa)t)} + d_m e^{i(m\phi - (m\Omega - \kappa)t)} \Big],\nonumber
\end{eqnarray}
where $a_m, b_m, c_m, d_m$ are amplitudes determined by the initial conditions, and $c_m = -\frac{i}{\gamma} a_m$, $d_m = \frac{i}{\gamma} b_m$.

\subsection{Small $m=2$ impact}\label{app:m2_impact}

The initial velocity kick by the perturber from Equation \ref{eq:m2_kick} with a simple $m=2$ mode with small amplitude can be written in compact, complex notation, as the real parts of  
\begin{equation}
\Delta V_R = \epsilon R e^{i2\phi}, \quad \Delta V_\phi = i\epsilon R e^{i2\phi}.
\end{equation}
For an $m=2$ mode, using the general solution (Eq.\,\ref{eq:fourier_solution_ap}) obtained in the previous section, we write
\begin{eqnarray}\label{eq:applied_ics}
\Delta V_R &=& a_2 e^{i(2\phi-(2\Omega + \kappa)t)} + b_2 e^{i(2\phi-(2\Omega - \kappa)t)},\\
\Delta V_\phi &=& c_2 e^{i(2\phi-(2\Omega + \kappa)t)} + d_2 e^{i(2\phi-(2\Omega - \kappa)t)}.\nonumber
\end{eqnarray}
At $t=0$, the solution must match the initial conditions, that is:  
\begin{eqnarray}
\Delta V_R(0) &=& (a_2+b_2)e^{i2\phi} = \epsilon R e^{i2\phi},\\
\Delta V_\phi(0) &=& (c_2+d_2)e^{i2\phi} = \frac{i}{\gamma} (-a_2 + b_2)e^{i2\phi} = i \epsilon R e^{i2\phi},
\end{eqnarray}
where we used the eigenmode relation $c_2 = -\frac{i}{\gamma} a_2$, $d_2 = \frac{i}{\gamma} b_2$. This immediately gives  
\begin{eqnarray}\label{eq:m2_resolution_linsys}
a_2 + b_2 &=& \epsilon R,\\
-a_2 + b_2 &=& \gamma \epsilon R.\nonumber
\end{eqnarray}
Solving these two simple equations for $a_2$ and $b_2$, we obtain the amplitude of the surviving modes
\begin{equation}\label{eq:params_simple}
a_2 = \epsilon R\frac{1 - \gamma}{2}, \quad b_2 = \epsilon R\frac{1 + \gamma}{2}.
\end{equation}
Substituting these amplitudes, along with the corresponding $c_2 = -\frac{i}{\gamma} a_2$ and $d_2 = \frac{i}{\gamma} b_2$, back into the general solution form (Eq.\,\ref{eq:applied_ics}) shows the time evolution of the perturbation:
\begin{eqnarray}\label{eq:applied_ics_resolved}
\Delta V_R&=& \epsilon R\frac{1 - \gamma}{2} e^{i(2\phi-(2\Omega + \kappa)t)} + \epsilon R\frac{1 + \gamma}{2} e^{i(2\phi-(2\Omega - \kappa)t)},\\
\Delta V_\phi &=& -\frac{i}{\gamma}\epsilon R\frac{1 - \gamma}{2} e^{i(2\phi-(2\Omega + \kappa)t)} + \frac{i}{\gamma}\epsilon R\frac{1 + \gamma}{2} e^{i(2\phi-(2\Omega - \kappa)t)}.\nonumber
\end{eqnarray}
These expressions clearly show that the initial impulsive condition at $t=0$ has decomposed into a superposition of two distinct waves, with amplitudes $a_2$ and $b_2$. Thus, the relative amplitude of these waves is directly determined by
\begin{equation}\label{eq:amplitudes_analytical_ap}
    \frac{\mathcal{A}_{\Omega - \kappa/2}}{\mathcal{A}_{\Omega + \kappa/2}} = \frac{|b_2|}{|a_2|} = \frac{1 + \gamma}{\gamma-1},
\end{equation}
where we used that $\gamma>1$ to remove the absolute values.

\section{Basis Function evolution}\label{app:basis_function}

The linear PDE system (Eq.\,\ref{eq:lin_syst_ap}) obtained in Sect.\,\ref{sect:analytic_derivation} (Appendix\,\ref{app:analytical_derivations}) governs the evolution of the small impact system. We can solve this PDE system efficiently by expressing the velocity perturbations in terms of coefficients from a basis function expansion. Using this method, the linear PDE system is transformed into a system of coupled ODEs for the time-dependent coefficients, allowing us to evolve the system ``only'' in the coefficient space.

\begin{figure*}
    \centering
    \includegraphics[width=0.98\linewidth]{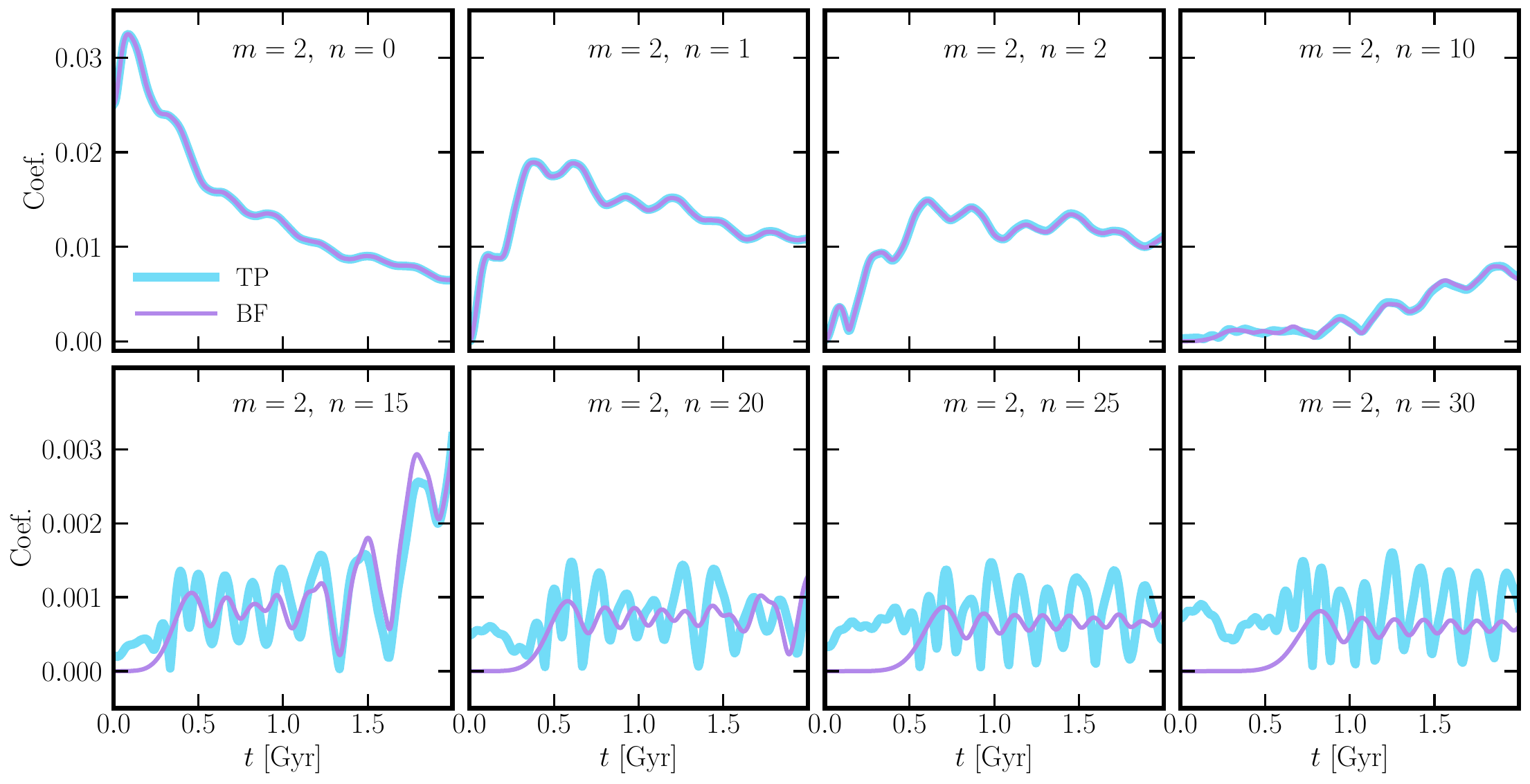}
    \caption{Temporal series of the basis function coefficients after a small $m=2$ impact (Sect.\,\ref{sect:simulations}). In blue, we show the basis function coefficient at each snapshot of the test particle simulation. In purple, we initialize the coefficients $\alpha_{m,n}(t=0)$ and $\beta_{m,n}(t=0)$ analytically and evolve them using Eq.\,\ref{eq:basis_function_regresion}.}
    \label{fig:basis_function}
\end{figure*}

We expand the perturbations $\Delta V_R(R,\phi)$ and $\Delta V_\phi(R,\phi)$ in an orthonormal basis function. For the angular direction, we use the full complex Fourier basis, while in the radial direction we adopt a general orthonormal basis $\{G_n(r)\}$ defined on the interval $R_{\min} \le r \le R_{\max}$. That is, we write
\begin{eqnarray}\label{eq:vrvphi_basis}
\Delta V_R(R,\phi) &=& \sum_{m,n} \alpha_{m,n}\, u_m(\phi)\, G_n(r)\\
\Delta V_\phi(R,\phi) &=& \sum_{m,n} \beta_{m,n}\, u_m(\phi)\, G_n(r),\nonumber
\end{eqnarray}
where the angular and radial basis functions are given by
\begin{equation}
u_m(\phi) = \frac{1}{\sqrt{2\pi}}\, e^{i m \phi},
\end{equation}
and the radial basis functions satisfy the orthonormality condition
\begin{equation}
\int_{R_{\min}}^{R_{\max}} r\, dr\, G_n(r)\, G_{n'}(r) = \delta_{n,n'}.
\end{equation}

Assuming that the quantities $\Delta V_R, \Delta V_\phi$ evolve in time according to Eq.\,\ref{eq:lin_syst_ap}, we substitute them by our basis functions equations (Eq.\,\ref{eq:vrvphi_basis}). Below we show an example of how we proceed with the $\Omega \frac{\partial \Delta V_R}{\partial \phi}$ term. Since it is a linear system, the calculation is equivalent for all the other terms of the system.

After inserting the basis expansion, the $\Omega \frac{\partial \Delta V_R}{\partial \phi}$ term takes the form
\begin{equation}
\Omega(r)\sum_{m,n} i\,m\,\alpha_{m,n}(t) \, u_{m}(\phi) \, G_{n}(r)
\end{equation}
where we have used that $\frac{\partial}{\partial \phi} u_m(\phi) = i\,m\,u_m(\phi)$.
To isolate the evolution of a particular coefficient $\alpha_{m,n}(t)$, we project this expression onto the basis by multiplying it by $u_m^*(\phi)\, G_n(r)$ and integrating over the entire domain
\begin{equation}
\int_0^{2\pi} d\phi \int r\, dr \, u_m^*(\phi)\, G_n(r) 
\left[ \sum_{m',n'} \alpha_{m',n'}(t)\, u_{m'}(\phi)\, \Omega(r)\, G_{n'}(r) \right].
\end{equation}
Due to the orthonormality of the angular basis functions $u_m(\phi)$, the expression simplifies to
\begin{equation}
\sum_{n'} \alpha_{m,n'}(t) \, \Omega_{n,n'},
\end{equation}
where the radial coupling integral is defined by
\begin{equation}\label{eq:radial_coupling}
\Omega_{n,n'} = \int r\, dr \, G_n(r)\, \Omega(r)\, G_{n'}(r).
\end{equation}

One can proceed identically with the rest of the terms in Eq.\,\ref{eq:lin_syst_ap}. Only the terms with $\partial/\partial\phi$ have the $i\,m$ factor. The derivation for the left part of the system, the temporal derivatives, leads to a coupling integral
\begin{equation}
I_{n,n'} = \int r\, dr \, G_n(r)\, G_{n'}(r).
\end{equation}
where the identity matrix appears due to the orthonormality of the radial basis.

After the projection of the entire system, the time evolution of the expansion coefficients corresponding to a given angular mode $m$ can be written as a system of coupled ODEs:
\begin{eqnarray}
\dot{\alpha}_{m,n} &=& - i\, m \sum_{n'} \Omega_{n,n'}\, \alpha_{m,n'} + \sum_{n'} (\kappa\gamma)_{n,n'}\, \beta_{m,n'},\\
\dot{\beta}_{m,n} &=& - i\, m \sum_{n'} \Omega_{n,n'}\, \beta_{m,n'} - \sum_{n'} \left(\frac{\kappa}{\gamma}\right)_{n,n'}\, \alpha_{m,n'},\nonumber
\end{eqnarray}
where $(\kappa\gamma)_{n,n'}$ and $(\kappa/\gamma)_{n,n'}$ are constructed using an analogue to Eq.\,\ref{eq:radial_coupling}. In compact block--matrix notation for each $m$, this system is expressed as
\begin{equation}\label{eq:basis_function_regresion}
\frac{d}{dt} \begin{pmatrix} \alpha_{m,n} \\ \beta_{m,n} \end{pmatrix} =   
\begin{pmatrix}
-i m \Omega_{n,n'} & (\kappa\gamma)_{n,n'} \\
\left(\frac{\kappa}{\gamma}\right)_{n,n'} & -i m \Omega_{n,n'} \end{pmatrix} 
\begin{pmatrix} \alpha_{m,n} \\ \beta_{m,n} \end{pmatrix}, 
\end{equation}
and can be solved numerically (e.g. using matrix exponentiation). This conversion of PDEs into ODEs using basis function expansion (BFE) is a core technique in galactic dynamics, often used in spectral methods and the response-matrix framework \citep[e.g.][]{kalnajs1971matrix,rozier2019matrix,rozier2022matrix_lmc,petersen2024linearresponse}. It transforms the differential operators in the PDE into algebraic operations on the time-dependent basis coefficients, allowing for efficient time evolution directly in coefficient space.

In Fig.\,\ref{fig:basis_function} we show the evolution of different $n$ coefficients of the orthogonal basis function for the test particle simulation (TP, blue lines). For this run, we used a Legendre polynomial basis in the radial direction. Using the analytical impact formula, we compute the initial $\alpha_{m,n},\,\beta_{m,n}$ coefficients, and evolve them using Eq.\,\ref{eq:basis_function_regresion} (BF, purple lines). We observe a great match between both methods in the low order coefficients. In the high order coefficients, we notice that we can go well below the noise threshold, and still capture a significant signal. This demonstrates the accuracy and efficiency of evolving the system in coefficient space for the linear case.

Applying this spectral approach to the SINDy-discovered non-linear PDE for the large-impact scenario (Eq.\,\ref{eq:nonlin_sys_sindy}) is more challenging than the linear case because non-linear terms introduce complex couplings between the basis coefficients. However, solving the resulting system of ODEs for the coefficients would still be significantly faster than running full particle simulations or solving the non-linear PDE numerically. This method could be applied to more complex SINDy-discovered PDEs, perhaps including self-gravity terms, offering an efficient way to explore the parameter space of galactic perturbations.

\section{Realistic Simulation Details}
\label{app:realistic_sim_details}

This appendix provides detailed information about the realistic simulations used in Section \ref{sect:n-body}. 

\paragraph{$N$-body simulation} We use the $N$-body simulation of a Sgr-like impact, presented in \citet{asano2025sim_sgr}, which is an extension of the original models described in \citet{fujii2019nbodyiso}. 
In this model, the MW-like host galaxy is constructed with a three-component structure: a DM halo, a classical bulge, and an exponential stellar disc. The DM halo follows an NFW profile \citep{navarro1996nfw} with a scale radius of $10\,$kpc and a mass of $M_{\text{h}} = 8.68\times10^{11}\,M_\odot$. The bulge is modelled with a \citet{hernquist1990halo} profile having a scale radius of $0.75\,$kpc and a mass of $5.42\times 10^{9}\,M_\odot$. The stellar disc is characterized by an exponential radial profile with a scale radius of $2.3\,$kpc, a vertical scale height of $0.2\,$kpc, and a total mass of $3.61\times10^{10}\,M_\odot$. The stellar disc contains $213$M particles. The Sgr-like satellite has a DM component that follows an NFW profile with a total mass of $5\times10^{10}\,M_\odot$ and a scale radius of $7.5\,$kpc, while its stellar component is represented by a Hernquist profile with a mass of $1\times10^{9}\,M_\odot$ and a scale radius of $2\,$kpc.

From this simulation, we select the time range starting $250\,$Myr after the first pericentre ($t=0.9\,$Gyr) for our analysis. This period resembles the tidal spiral phase before the appearance of kinematic crossings (Eq.\,\ref{eq:shock}) becomes significant.

\paragraph{Test Particle Rerun \texttt{TPR}} Complementing this $N$-body simulation, Asano et al. (2025, in prep.) have run a suite of test particle reruns of the previous $N$-body model using \texttt{Agama}. Here we used one of them, which we call \texttt{TPR}, to isolate and test the contribution of self-gravity to the SINDy analysis compared to a frozen potential. In this run, the initial conditions (ICs) of the disc particles are those of the $N$-body at the first pericentre time ($t=0.9\,$Gyr). The particles are then evolved in the isolated frozen axisymmetric potential of the MW host galaxy, effectively removing the self-gravity of the disc and halo, as well as the external acceleration from the perturber. The evolution of this simplified model is shown alongside the $N$-body maps in the second and fourth columns of Fig.\,\ref{fig:nbody_velocity_maps}.

\end{document}